\documentclass[sigconf]{acmart}

\usepackage{amsmath,bm}
\usepackage{enumitem}
\usepackage{xspace}
\usepackage{graphicx}
\usepackage{subfigure}
\usepackage{multirow}
\usepackage{balance}
\AtBeginDocument{%
  \providecommand\BibTeX{{%
    \normalfont B\kern-0.5em{\scshape i\kern-0.25em b}\kern-0.8em\TeX}}}

\setcopyright{acmcopyright}

\newcommand{\model}{AdaGCL\xspace}

\copyrightyear{2023}
\acmYear{2023}
\setcopyright{acmlicensed}\acmConference[KDD '23]{Proceedings of the 29th ACM SIGKDD Conference on Knowledge Discovery and Data Mining}{August 6--10, 2023}{Long Beach, CA, USA}
\acmBooktitle{Proceedings of the 29th ACM SIGKDD Conference on Knowledge Discovery and Data Mining (KDD '23), August 6--10, 2023, Long Beach, CA, USA}
\acmPrice{15.00}
\acmDOI{10.1145/3580305.3599768}
\acmISBN{979-8-4007-0103-0/23/08}

\begin{document}



\title{Adaptive Graph Contrastive Learning for Recommendation}




\author{Yangqin Jiang}
\affiliation{%
  \institution{University of Hong Kong}
  \city{Hong Kong}
  \country{China}
}
\email{mrjiangyq99@gmail.com}

\author{Chao Huang}
\authornote{Chao Huang is the corresponding author.}
\affiliation{%
  \institution{University of Hong Kong}
  \city{Hong Kong}
  \country{China}
}
\email{chaohuang75@gmail.com}

\author{Lianghao Xia}
\affiliation{%
  \institution{University of Hong Kong}
  \city{Hong Kong}
  \country{China}
}
\email{aka_xia@foxmail.com}


\begin{abstract}

Graph neural networks (GNNs) have recently emerged as an effective collaborative filtering (CF) approaches for recommender systems. The key idea of GNN-based recommender systems is to recursively perform message passing along user-item interaction edges to refine encoded embeddings, relying on sufficient and high-quality training data. However, user behavior data in practical recommendation scenarios is often noisy and exhibits skewed distribution. To address these issues, some recommendation approaches, such as SGL, leverage self-supervised learning to improve user representations. These approaches conduct self-supervised learning through creating contrastive views, but they depend on the tedious trial-and-error selection of augmentation methods. In this paper, we propose a novel \underline{Ada}ptive \underline{G}raph \underline{C}ontrastive \underline{L}earning (\model) framework that conducts data augmentation with two adaptive contrastive view generators to better empower the CF paradigm. Specifically, we use two trainable view generators - a graph generative model and a graph denoising model - to create adaptive contrastive views. With two adaptive contrastive views, \model\ introduces additional high-quality training signals into the CF paradigm, helping to alleviate data sparsity and noise issues. Extensive experiments on three real-world datasets demonstrate the superiority of our model over various state-of-the-art recommendation methods. Our model implementation codes are available at the link \url{https://github.com/HKUDS/AdaGCL}.


\end{abstract}



\maketitle

\section{Introduction}


Recommender systems are a crucial tool for web applications, helping users to navigate the overwhelming amount of information available online. These systems provide personalized recommendations of items that users might be interested in, such as products on online retail platforms~\cite{wu2018turning,wang2020time}, posts on social networking sites~\cite{zhang2021understanding,jamali2010matrix}, and video sharing platforms~\cite{zhan2022deconfounding,2023mmssl}. One of the most common approaches for generating these recommendations is collaborative filtering (CF), where the system uses the preferences of similar users or items to suggest new items for a given user~\cite{he2017neural,xia2023graph}.


Collaborative filtering (CF) models have traditionally relied on matrix factorization (MF) to learn latent user and item embeddings from interaction data. However, with the rise of graph neural networks (GNNs), there has been a growing interest in using these models to propagate information along the user-item interaction graph and learn more sophisticated representations of user-item interactions. PinSage~\cite{ying2018graph}, NGCF~\cite{wang2019neural}, and LightGCN~\cite{he2020lightgcn} are examples of GNN-based CF models that have shown promising results in personalized recommendations. These models use graph convolutional networks (GCNs) to propagate embeddings over the user-item interaction graph, allowing them to capture higher-order interactions between users and items that are not captured by other alternative CF models. In particular, PinSage and NGCF use multi-layer GCNs to capture both local and global information about the user-item interaction graph, while LightGCN simplifies the message passing process by omitting the non-linear transformer and only using a simple weighted sum of the neighboring embeddings.

Graph-based collaborative filtering models have become increasingly popular in recommender systems. However, these models face challenges that current techniques have not adequately addressed. One such challenge is data noise, which can arise due to various factors, such as users clicking on irrelevant products due to over-recommendation of popular items. Directly aggregating information from all interaction edges in the user-item interaction graph can lead to inaccuracies in user representations, and multi-hop embedding propagation can worsen the noise effect. Therefore, existing graph-based CF models may not accurately capture user interests and generate inaccurate recommendations. Furthermore, the sparsity and skewed distribution of recommendation data can negatively impact effective user-item interaction modeling. As a result, current approaches may suffer from the problem of user data scarcity, where high-quality training signals may be limited.

Recently, some recommendation methods, such as SGL~\cite{wu2021self}, SLRec~\cite{yao2021self} and HCCF~\cite{xia2022hypergraph}, have leveraged self-supervised learning to improve user representations. These methods introduce additional supervision information by creating contrastive views through probability-based random masking or adding noise. However, these operations may keep some noisy interactions or drop important training signals during the data augmentation process, limiting the applicability and potential of contrastive learning. \\\vspace{-0.1in}


\noindent \textbf{Contribution.} Given the limitations and challenges of existing solutions, we propose a novel Adaptive Graph Contrastive Learning (\model) framework to enhance the robustness and generalization performance of recommender systems. Our approach leverages adaptive contrastive learning to introduce high-quality training signals, empowering the graph neural CF paradigm. While several recent studies have used contrastive learning to improve model performance, they all require specific ways to create contrastive views. The selection of methods for creating contrastive views can be burdensome and often limited to a pool of prefabricated views, which can limit their potential and applicability. To address these issues, we integrate a graph generative model and a graph denoising model to establish views that adapt to the data distribution, achieving adaptive contrastive views for graph contrastive learning. By providing two different and adaptive views, we offer additional high-quality training signals that can enhance the graph neural CF paradigm and help address the problem of model collapse in contrastive learning-based data augmentation.


In summary, this paper makes the following contributions: \vspace{-0.05in}
\begin{itemize}[leftmargin=*]
    \item We propose a novel self-supervised recommendation model, called \model, that enhances the robustness of the graph CF by distilling additional training signals from adaptive contrastive learning. \\\vspace{-0.1in}
    

    \item \model\ employs two trainable view generators, namely a graph generator and a graph denoising model, to create contrastive views. These views address the problem of model collapse and enable adaptive views for contrastive learning, ultimately enhancing the effectiveness of the graph neural CF paradigm. \\\vspace{-0.1in}
    

    \item Our experimental results demonstrate that our \model\ outperforms various baseline models on multiple datasets, highlighting its superior performance and effectiveness. Furthermore, our approach is able to address the challenges of data noise and user data scarcity, which can negatively impact the accuracy of collaborative filtering models for recommendation.
    
\end{itemize}

\begin{figure*}[t]
    \centering
    \includegraphics[width=1 \linewidth]{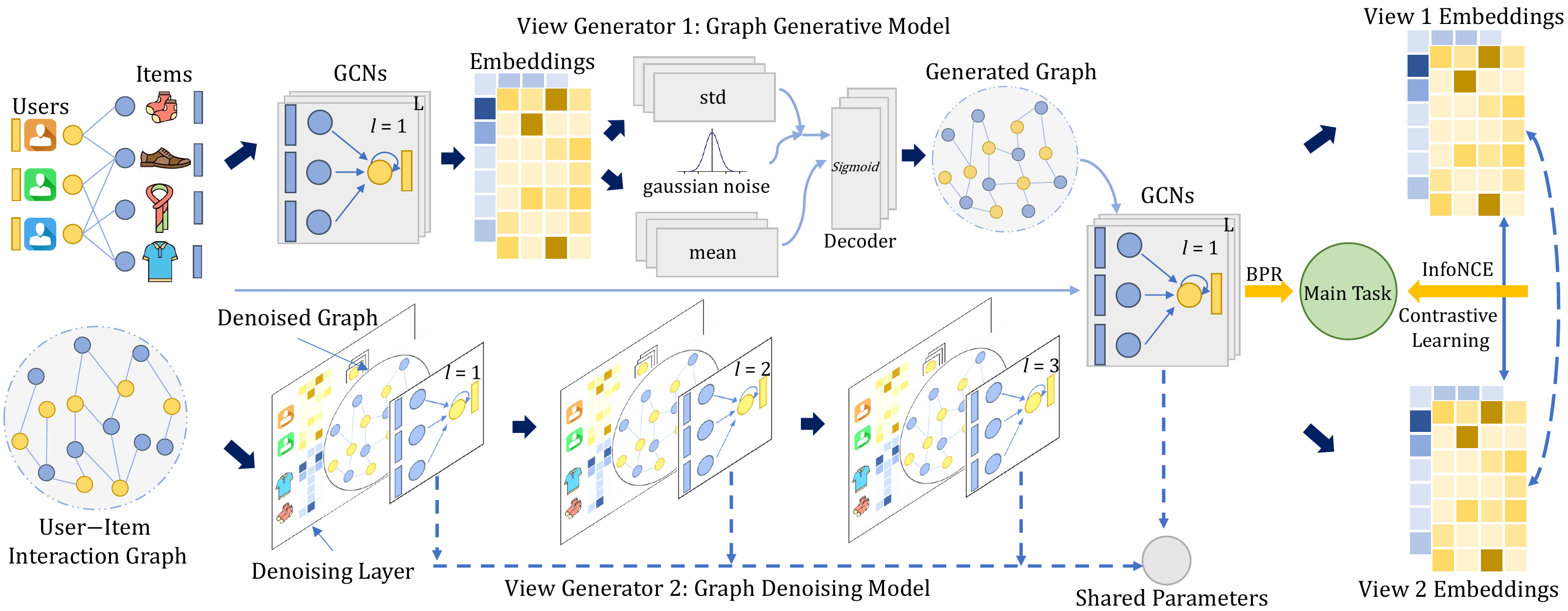}
    \vspace{-0.15in}
    \caption{Overall framework of the proposed \model model.}
    \label{fig:figure_overall}
\end{figure*}

\section{PRELIMINARIES AND RELATED WORK}

\subsection{Collaborative Filtering Paradigm}

We let $\mathcal{U}$ $=$ \{$u_1, \cdots, u_i, $ $\cdots, u_{I}$\} ($|\mathcal{U}|=I$) and $\mathcal{V}$ $=$ \{$v_1, \cdots, v_j, \cdots, v_{J}$\} ($|\mathcal{V}|=J$) represent the set of users and items, respectively. The interaction matrix $\mathcal{A} \in \mathbb{R}^{\mathcal{I} \times \mathcal{J}}$ indicates the implicit relationships between each user in $\mathcal{U}$ and his/her consumed items. Each entry $\mathcal{A}_{i,j}$ in $\mathcal{A}$ will be set as $1$ if user $u_i$ has adopted item $v_j$ before and $\mathcal{A}_{i,j}=0$ otherwise. Upon the constructed interaction graph structures, the core component of graph-based CF paradigm lies in the information aggregation function, gathering the feature embeddings of neighboring users/items via different aggregators, e.g., mean or sum. The objective of CF task is to forecast the unobserved user-item interactions with the encoded corresponding representations. The assumption of the collaborative filtering paradigm is that users who exhibit similar behavior are more likely to share similar interests. One popular paradigm of existing collaborative filtering (CF) approaches involves using various embedding functions to generate vectorized representations of users and items. The similarity matching function is then introduced to estimate the relevance score between a user $u_i$ and a candidate item $v_j$.


 \subsection{Graph-based Recommender Systems}


Graph neural architectures have become increasingly popular in recent years due to their ability to effectively model complex relationships between users and items in recommendation systems~\cite{wu2022graph}. These architectures leverage graph embedding propagation techniques to encode the interactions between users and items in the form of graph embeddings. One important advantage of graph neural architectures is their ability to capture multi-hop connections between users and items. This allows the model to capture more complex and nuanced relationships between users and items. Some architectures, like PinSage~\cite{ying2018graph} and NGCF~\cite{wang2019neural}, use graph convolutional networks in the spectral domain. Others, like LightGCN~\cite{he2020lightgcn}, simplify the non-linear transformation and use sum-based pooling over neighboring representations for improved efficiency. These architectures encode each user and item into transformed embeddings while preserving multi-hop connections. 

Moreover, fine-grained graph-based relational learning techniques among users and items have been introduced in graph neural networks for user/item representations. Examples of these techniques include DGCF~\cite{wang2020disentangled}, DCCF~\cite{ren2023disentangled}, and DRAN~\cite{wang2022learning}. These techniques aim to learn disentangled or behavior-aware user representations by exploiting the graph-structured multi-intent information. In addition to these techniques, graph neural networks have also been increasingly used in next-item recommendation tasks to capture the temporal dependencies between items and how a user's preferences for certain items evolve over time. Models such as DGSR~\cite{zhang2022dynamic}, RetaGNN~\cite{hsu2021retagnn}, and GCE-GNN~\cite{wang2020global} represent the user's historical interactions as a sequence of items and use graph-based message passing to update each item's embedding based on the information from its neighbors. This approach allows the models to capture the dependencies and relationships between items and how a user's preferences for certain items evolve over time, leading to more accurate and effective recommendations.

\subsection{Self-Supervised Graph Learning}

Despite the success of supervised learning in many applications, obtaining a large labeled dataset can be a challenging and expensive task. To overcome this limitation, self-supervised learning (SSL) has emerged as a promising solution. In the context of graph machine learning, SSL has been shown to be effective for learning high-quality representations of graph data. One of the recent advances in SSL is the use of contrastive learning with auxiliary training signals generated from various graph data, such as heterogeneous graph~\cite{hwang2020self}, spatio-temporal graph~\cite{zhang2023automated} and molecular graph~\cite{zhang2021motif}. SSL with contrastive learning has been shown to improve the quality of embeddings for graphs, leading to better performance on tasks, such as node classification and link prediction. 

Self-supervised graph learning has also been introduced into recommender systems, to show great potential for enhancing representations of users and items with contrastive SSL~\cite{wu2021self} or generative SSL~\cite{li2023graph} techniques. One example of a self-supervised graph learning framework is SGL~\cite{wu2021self}, which generates contrastive views of the user-item interaction graph using random node and edge dropout operations. By maximizing the agreement between the embeddings of the contrastive views, SSL signals can be incorporated into the model joint learning process. Another example is GFormer~\cite{li2023graph}, which leverages the graph autoencoder to reconstruct the masked user-item interactions for augmentation. By generating augmented training data in this way, the model can learn more effective representations of users and items. Additionally, the use of self-supervised graph learning techniques has benefited a variety of recommendation scenarios. For example, S3-Rec~\cite{zhou2020s3} S3-Rec is based on a self-attentive neural architecture and uses four auxiliary self-supervised objectives to learn the correlations among various types of data, including attributes, items, subsequences. C2DSR~\cite{cao2022contrastive} is a cross-domain sequential recommendation approach that proposes a contrastive cross-domain infomax objective to enhance the correlation between single- and cross-domain user representations. SLMRec~\cite{tao2022self} is a SSL approach for multimedia recommendation that captures multi-modal patterns in the data.

\section{METHODOLOGY}

In this section, we introduce the \model\ framework, which is composed of three parts. The first part uses a graph message passing encoder to capture local collaborative relationships among users and items. The second part proposes a novel adaptive self-supervised learning framework that includes two trainable view generators made of variational and denoising graph models. The third part introduces the phase of model optimization. The overall architecture of the \model\ model is illustrated in Figure~\ref{fig:figure_overall}.

\subsection{Local Collaborative Relation Learning}


To encode the interaction patterns between users and items, we follow the common collaborative filtering paradigm by embedding them into a $d$-dimensional latent space. Specifically, we generate embedding vectors $\mathbf{e}_i$ and $\mathbf{e}_j$ of size $\mathbb{R}^{d}$ for user $u_i$ and item $v_j$, respectively. We also define embedding matrices $\mathbf{E}^{(u)} \in \mathbb{R}^{I\times d}$ and $\mathbf{E}^{(v)} \in \mathbb{R}^{J\times d}$ to represent the embeddings of users and items, respectively. To propagate the embeddings, we design a local graph embedding propagation layer inspired by the simplified graph convolutional network used in LightGCN~\cite{he2020lightgcn}.
\begin{align}
    \mathbf{z}_{i}^{(u)} = \bar{\mathcal{A}}_{i,*} \cdot \mathbf{E}^{(v)}, \ \ \ \ \ \mathbf{z}_{j}^{(v)} = \bar{\mathcal{A}}_{*,j} \cdot \mathbf{E}^{(u)},
\end{align}
\noindent To represent the aggregated information from neighboring items/users to the central node $u_i$ and $v_j$, we use the vectors $\mathbf{z}{i}^{(u)}$ and $\mathbf{z}{j}^{(v)}$ respectively, both having a dimension of $\mathbb{R}^{d}$. We derive the normalized adjacent matrix $\bar{\mathcal{A}}\in \mathbb{R}^{I\times J}$ from the user-item interaction matrix $\mathcal{A}$. Specifically, $\bar{\mathcal{A}}$ is calculated using the following formula:

\begin{equation}
    \bar{\mathcal{A}} = \mathbf{D}_{(u)}^{-1/2}\cdot \mathcal{A} \cdot \mathbf{D}_{(v)}^{-1/2}, \ \ \ \ \bar{\mathcal{A}}_{i,j} = \frac{\mathcal{A}_{i,j}}{\sqrt{|\mathcal{N}_{i}|\cdot |\mathcal{N}_{j}|}},
\end{equation}
\noindent The diagonal degree matrices for users and items are $\mathbf{D}{(u)} \in \mathbb{R}^{I\times I}$ and $\mathbf{D}{(v)} \in \mathbb{R}^{J\times J}$ respectively. The neighboring items/users of user $u_i$ and item $v_j$ are denoted by $\mathcal{N}_i$ and $\mathcal{N}_j$ respectively.


To refine the user/item representations and aggregate local neighborhood information for contextual embeddings, we integrate multiple embedding propagation layers. We denote the embedding of user $u_i$ and item $v_j$ at the $l$-th graph neural network (GNN) layer as $\mathbf{e}{i,l}^{(u)}$ and $\mathbf{e}{j,l}^{(v)}$ respectively. We formally define the message passing process from the ($l-1$)-th layer to the $l$-th layer as follows:
\begin{equation}
    \mathbf{e}_{i,l}^{(u)} = \mathbf{z}_{i,l}^{(u)} + \mathbf{e}_{i,l-1}^{(u)}, \ \ \ \mathbf{e}_{j,l}^{(v)} = \mathbf{z}_{j,l}^{(v)} + \mathbf{e}_{j,l-1}^{(v)}.
\end{equation}
To obtain the embedding for a node, we sum its embeddings across all layers. The inner product between the final embedding of a user $u_i$ and an item $v_j$ is used to predict $u_i$'s preference towards $v_j$:
\begin{equation}
    \mathbf{e}_{i}^{(u)} = \sum_{l=0}^{L}\mathbf{e}_{i,l}^{(u)},\ \ \mathbf{e}_{j}^{(v)} = \sum_{l=0}^{L}\mathbf{e}_{j,l}^{(v)},\ \  \hat{y}_{i,j} = \mathbf{e}_{i}^{(u)\top}\mathbf{e}_{j}^{(v)}.
\end{equation}

\subsection{Adaptive View Generators for\\Graph Contrastive Learning}
\subsubsection{\bf Dual-View GCL Paradigm}

Existing graph contrastive learning (GCL) methods, such as those proposed in \cite{wu2021self, xia2022hypergraph, lin2022improving}, generate views in specific ways, such as randomly dropping edges, nodes, or constructing hypergraphs. However, selecting an appropriate method for generating views can be burdensome, as it often relies on tedious trial-and-error or a limited pool of prefabricated views. This limitation can restrict the applicability and potential of these methods. To overcome this issue, we propose using two learnable view generators to obtain adaptive views for GCL.


Developing view generators for graph contrastive learning methods poses a challenge due to the risk of model collapse, where two views generated by the same generator share the same distribution, potentially leading to inaccurate contrastive optimization. To address this challenge, we propose using two distinct view generators that augment the user-item graph from different perspectives. Specifically, we employ a graph generative model and a graph denoising model as our two view generators. The graph generative model is responsible for reconstructing views based on graph distributions, while the graph denoising model leverages the graph's topological information to remove noise from the user-item graph and generate a new view with less noise.

In line with existing self-supervised collaborative filtering (CF) paradigms, such as those proposed in \cite{wu2021self,xia2022hypergraph}, we use node self-discrimination to generate positive and negative pairs. Specifically, we treat the views of the same node as positive pairs (i.e., {($\textbf{e}'{i}$, $\textbf{e}''{i}$)|$u_i \in \mathcal{U}$}), and the views of any two different nodes as negative pairs (i.e., {($\textbf{e}'{i}$, $\textbf{e}''{i'}$)|$u_i,u_{i'} \in \mathcal{U}$, $u_i \neq u_{i'}$}). Formally, the contrastive loss function that maximizes the agreement of positive pairs and minimizes that of negative pairs is as follows:
\begin{equation}
    \label{eq:infoNCE}
    \mathcal{L}_{ssl}^{user} = \sum_{u_i\in\mathcal{U}} -\text{log} \frac{\text{exp}(s(\textbf{e}'_{i}, \textbf{e}''_{i})/\tau)}{\sum_{u_{i'} \in \mathcal{U}}\text{exp}(s(\textbf{e}'_{i},\textbf{e}''_{i'}/\tau)},
\end{equation}
To measure the similarity between two vectors, we use the cosine similarity function denoted by $s(\cdot)$, with the hyper-parameter $\tau$ known as the \textit{temperature} in softmax. We compute the contrastive loss for the item side as $\mathcal{L}_{ssl}^{item}$ in a similar way. By combining these two losses, we obtain the objective function for the self-supervised task, which is denoted by $\mathcal{L}_{ssl} = \mathcal{L}_{ssl}^{user} + \mathcal{L}_{ssl}^{item}$.

\begin{figure}[t]
    \centering
    \includegraphics[width=1 \linewidth]{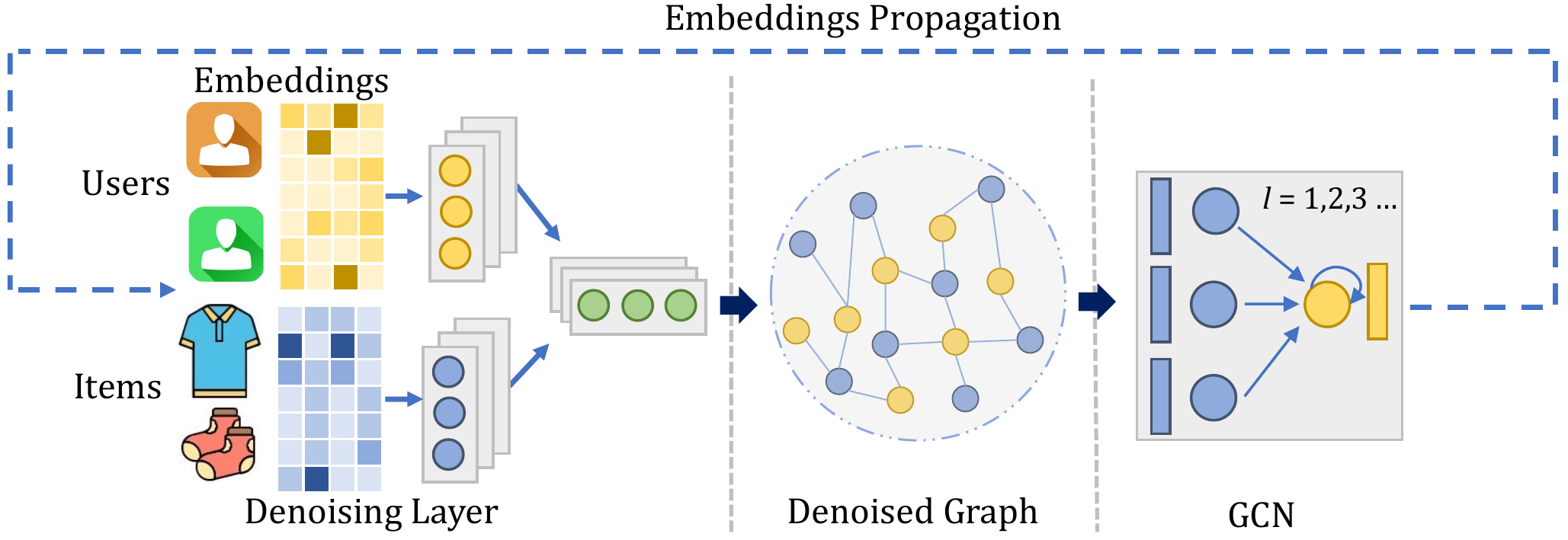}
    \vspace{-0.1in}
    \caption{Workflow of the graph denoising model.}
    \label{fig:figure_denoise}
\end{figure}

\subsubsection{\bf Graph Generative Model as View Generator}

The recent emergence of learning-based graph generative models~\cite{kipf2016variational, wang2019learning} provides a promising solution for view generator. In this study, we adopt the widely-used Variational Graph Auto-Encoder (VGAE)~\cite{kipf2016variational} as the generative model, which combines the concept of variational auto-encoder with graph generation. Compared to GAE, VGAE incorporates KL divergence to reduce the risk of overfitting, allowing for more diverse graphs to be generated by increasing the uncertainty. This feature provides a more challenging contrastive view for contrastive learning. Additionally, VGAE is relatively easier to train and faster than other currently popular generation models such as generative adversarial networks and diffusion models.

As illustrated in Fig.~\ref{fig:figure_overall}, we utilize a multi-layer GCN as the encoder to obtain the graph embeddings. Two MLPs are utilized to derive the mean value and the standard deviation of the graph embedding, respectively. With another MLP as the decoder, the input mean value and the standard deviation with Gaussian noise will be decoded to generate a new graph. The loss of VGAE is defined:

\begin{equation}
    \mathcal{L}_{gen} = \mathcal{L}_{kl} + \mathcal{L}_{dis},
\end{equation}
The term $\mathcal{L}_{kl}$ refers to the Kullback–Leibler divergence (KL divergence) between the distribution of node embeddings and the standard Gaussian distribution. On the other hand, $\mathcal{L}_{dis}$ is a cross-entropy loss that quantifies the dissimilarities between the generated graph and the original graph.

\subsubsection{\bf Graph Denoising Model as View Generator}
%

GNN models use message passing mechanisms to propagate and aggregate information along the input graph to learn node representations. However, the quality of the input graph can heavily impact model performance since messages aggregated along noisy edges can decrease the quality of node embeddings. Therefore, for the second view generator, we aim to generate a denoising view that can enhance model performance against noisy data.


To improve the quality of node embeddings obtained after each layer of GCN, we propose a graph neural network that incorporates a denoising layer to filter out noisy edges in the input graph. This parameterized network is shown in Fig.~\ref{fig:figure_denoise}. 
The main concept behind our approach is to actively filter out noisy edges in the input graph using a parameterized network.
For the $l$-th GCN layer, we use a binary matrix $\mathbf{M}^{l} \in {0,1}^{|\mathcal{V}|\times|\mathcal{V}|}$, where $m_{i,j}^{l}$ denotes whether the edge between node $u_i$ and $v_j$ is present ($0$ indicates a noisy edge).

Formally, the adjacency matrix of the resulting subgraph is $\mathbf{A}^{l} = \mathbf{A} \odot \mathbf{M}^{l}$, where $\odot$ is the element-wise product. 
The straightforward idea to reduce noisy edges with the least assumptions about $\mathbf{A}^{l}$ is to penalize the number of non-zero entries in $\mathbf{M}^{l}$ of different layers.

\begin{equation}
\label{eq:denoise}
    \sum_{l=1}^{L}{||\mathbf{M}^{l}||}_{0} = \sum_{l=1}^{L} \sum_{(u,v)\in \varepsilon} \mathbb{I}[m_{i,j}^{l} \neq 0],
\end{equation}
where $\mathbb{I}[\cdot]$ is an indicator function, with $\mathbb{I}[True] = 1$ and $\mathbb{I}[False]=0$, ${||\cdot||}_{0}$ represents the $l_{0}$ norm.
However, because of its combinatorial and non-differentiability nature, optimizing this penalty is computationally intractable.
Therefore, we consider each binary number $m_{i,j}^{l}$ to be drawn from a Bernoulli distribution parameterized by $\pi_{i,j}^{l}$, i.e., $m_{i,j}^{l} \sim \text{Bern}(\pi_{i,j}^{l})$.
Here, $\pi_{i,j}^{l}$ describes the quality of the edge ($u$, $v$).
%
To efficiently optimize subgraphs with gradient methods, we adopt the reparameterization trick and relax the binary entries $m_{i,j}^{l}$ from being drawn from a Bernoulli distribution to a deterministic function $g$ of parameters $\alpha_{i,j}^{l} \in \mathbb{R}$ and an independent random variable $\varepsilon^{l}$.
That is $m_{i,j}^{l} = g(\alpha_{i,j}^{l}, \varepsilon^{l})$.
%

Based on above operations, we design a denoising layer to learn the parameter $\alpha_{i,j}^{l}$ that controls whether to remove the edge ($u$, $v$).
 For the $l$-th GNN layer, we calculate $\alpha_{i,j}^{l}$ for user node $u$ and its interacted item node $v$ with $\alpha_{i,j}^{l} = f_{\theta^{l}}^{l}(\mathbf{e}_{i}^{l}, \mathbf{e}_{j}^{l})$, where $f_{\theta^{l}}^{l}$ is an MLP parameterized by $\theta^{l}$.
 In order to get $m_{i,j}^{l}$, we also utilize the concrete distribution along with a hard sigmoid function.
 Within the above formulation, the constraint on the number of non-zero entries in $\mathbf{M}^{l}$ in Eq.~(\ref{eq:denoise}) can be reformulated with:
\begin{equation}
    \mathcal{L}_{c} = \sum_{l=1}^{L} \sum_{(u_i,v_j)\in\varepsilon}(1 - \mathbb{P}_{\sigma(s_{i,j}^{l})}(0|\theta^{l})),
\end{equation}
 where $\mathbb{P}_{\sigma(s_{i,j}^{l})}$ is the cumulative distribution function (CDF) of $\sigma(s_{i,j}^{l})$, $\sigma(\cdot)$ extends the range of $s_{i,j}^{l}$, and $s_{i,j}^{l}$ is drawn from a binary concrete distribution with $\alpha_{i,j}^{l}$ parameterizing the location.

\subsection{Learning Task-aware View Generators}
Although two view generators could learn to generate better views from different aspects, there may be no optimization signals to adjust generated views to the main CF task.
The straightforward idea is introducing commonly-used BPR loss, as follows:
\begin{equation}
    \label{eq:bpr}
    \mathcal{L}_{bpr} = \sum_{(u,i,j)\in\mathcal{O}} - \text{log}\sigma(\hat{y}_{ui}-\hat{y}_{uj}),
\end{equation}
\noindent The training data is represented by $\mathcal{O} = (u,i,j)|(u,i) \in \mathcal{O}^{+}, (u,j)$ $\in \mathcal{O}^{-}$, where $\mathcal{O}^{+}$ denotes the observed interactions and $\mathcal{O}^{-} = \mathcal{U} \times \mathcal{I} / \mathcal{O}^{+}$ denotes the unobserved interactions.

To train the graph generative model, we use the node embeddings encoded by the VGAE encoder to compute BPR loss. The loss function $\mathcal{L}_{gen}$ is then updated as follows:
\begin{equation}
\label{eq:loss_gen}
    \mathcal{L}_{gen} = \mathcal{L}_{kl} + \mathcal{L}_{dis} + \mathcal{L}_{bpr}^{gen} + \lambda_2 ||\Theta||_{\text{F}}^{2},
\end{equation}
where $\Theta$ is the set of model parameters, while $\lambda_2$ is a hyperparameter used to control the strength of the weight-decay regularization.

To train the graph denoising model, we use the node embeddings obtained by the denoising neural network to compute the BPR loss. The loss function $\mathcal{L}_{den}$ is updated as follows:
\begin{equation}
\label{eq:loss_den}
    \mathcal{L}_{den} = \mathcal{L}_{c} + \mathcal{L}_{bpr}^{den} + \lambda_2 ||\Theta||_{\text{F}}^{2}.
\end{equation}


\subsection{Model Training}

The training of our proposed model consists of two parts.
In the upper-level part, we adopt a multi-task training strategy to jointly optimize the classic recommendation task (Eq.~(\ref{eq:bpr})) and the self-supervised learning task (Eq.~(\ref{eq:infoNCE})):
\begin{equation}
    \mathcal{L}_{upper} = \mathcal{L}_{bpr} + \lambda_1 \mathcal{L}_{ssl} + \lambda_2||\Theta||_{\text{F}}^{2},
\end{equation}
where $\Theta$ refers to the set of model parameters in the main task, which in this work, is the set of parameters of LightGCN. Additionally, $\lambda_1$ and $\lambda_2$ are hyperparameters that control the strengths of SSL and $L_2$ regularization, respectively. 

The lower-level part of the training involves optimizing the generative and denoising view generators based on Eq.~(\ref{eq:loss_gen}) and Eq.~(\ref{eq:loss_den}), which is formally presented as follows:
%
\begin{equation}
    \mathcal{L}_{lower} = \mathcal{L}_{gen} + \mathcal{L}_{den}.
\end{equation}

\subsection{Time Complexity Analysis}

We analyze the time complexity of our proposed model by considering its three key components. Firstly, the local collaborative relation learning module takes $O(L \times |\mathcal{A}| \times d)$ time, which is the same as that of LightGCN. Here, $L$ denotes the number of graph neural layers, $|\mathcal{A}|$ is the number of edges in the user-item interaction graph, and $d$ denotes the embedding dimensionality. Secondly, the graph generative model (VGAE) costs $O(|\mathcal{A}| \times d^2)$ time. Thirdly, the denoising layers in the graph denoising model cost $O(L \times |\mathcal{A}| \times d^2)$ time. Finally, the contrastive learning paradigm costs $O(L \times B \times (I + J) \times d)$, where $B$ denotes the number of users/items included in a single batch. $I$ and $J$ denote the number of users and items, respectively.

\section{EVALUATION}

To evaluate the effectiveness of our proposed model, our experiments are designed to answer the following research questions:
\begin{itemize}[leftmargin=*]
	\item \textbf{RQ1}: What is the performance of our proposed model compared to various state-of-the-art recommender systems? \\\vspace{-0.1in}
 
    \item \textbf{RQ2}: How do the key components of our proposed model contribute to its overall performance on different datasets? \\\vspace{-0.1in}
    
    \item \textbf{RQ3}: How well can our proposed model handle noisy and sparse data compared to baseline methods? \\\vspace{-0.1in}
    
    \item \textbf{RQ4}: How do the key hyperparameters influence the performance of our proposed model framework?
    
\end{itemize}

\begin{table}[t]
  \caption{Statistics of the experimental datasets.}
  \label{tab:dataset}
  \vspace{-0.1in}
  \small
  \centering
  \begin{tabular}{c|c|c|c|c}
    \hline
    \textbf{Dataset} & \textbf{User} $\#$ & \textbf{Item} $\#$ & \textbf{Interaction} $\#$ & \textbf{Density}\\
    \hline
    \hline
    Last.FM & 1,892 & 17,632 & 92,834 & $2.8\times10^{-3}$\\
    \hline
    Yelp & 42,712 & 26,822 & 182,357 & $1.6\times10^{-4}$\\
    \hline
    BeerAdvocate & 10,456 & 13,845 & 1,381,094 & $9.5\times10^{-3}$\\
    \hline
  \end{tabular}
  \vspace{-0.1in}
\end{table}

\begin{table*}[!htbp]
    \centering
    \caption{Performance comparison on Last.FM, Yelp, BeerAdvocate datasets in terms of \textit{Recall} and \textit{NDCG}.}
    \vspace{-0.1in}
    \resizebox{\linewidth}{!}{
    \begin{tabular}{|c|c|c|c|c|c|c|c|c|c|c|c|c|c|c|c|c|c|c|}
        \hline
        \textbf{Dataset} & \textbf{Metric} & 
        \textbf{BiasMF} & \textbf{NCF} & \textbf{AutoR} & \textbf{PinSage} &\textbf{STGCN} &
        \textbf{GCMC} & \textbf{NGCF} & \textbf{GCCF} & \textbf{LightGCN} & \textbf{SLRec} &
        \textbf{NCL} & \textbf{SGL} & \textbf{HCCF} & \textbf{SHT} & \textbf{DirectAU} & \textbf{Ours} & \textbf{p-val.}\\
        \hline
        \hline
        \multirow{4}{*}{Last.FM} & Recall@20 & 0.1879 & 0.1130 & 0.1518 & 0.1690 & 0.2067 & 0.2218 & 0.2081 & 0.2222 & 0.2349 & 0.1957 & 0.2353 & 0.2427 & 0.2410 & 0.2420 & 0.2422 & \textbf{0.2603} & 2.1$e^{-5}$ \\
        & NDCG@20 & 0.1362 & 0.0795 & 0.1114 & 0.1228 & 0.1528 & 0.1558 & 0.1474 & 0.1642 & 0.1704 & 0.1442 & 0.1715 & 0.1761 & 0.1773 & 0.1770 & 0.1727 & \textbf{0.1911} & 9.5$e^{-5}$ \\
        \cline{2-19}
        & Recall@40 & 0.2660 & 0.1693 & 0.2174 & 0.2402 & 0.2940 & 0.3149 & 0.2944 & 0.3083 & 0.3220 & 0.2792 & 0.3252 & 0.3405 & 0.3232 & 0.3235 & 0.3356 & \textbf{0.3531} & 6.9$e^{-5}$ \\
        & NDCG@40 & 0.1653 & 0.0952 & 0.1336 & 0.1472 & 0.1821 & 0.1897 & 0.1829 & 0.1931 & 0.2022 & 0.1737 & 0.2033 & 0.2104 & 0.2051 & 0.2055 & 0.2042 & \textbf{0.2204} & 5.6$e^{-4}$ \\
        \hline
        \multirow{4}{*}{Yelp} & Recall@20 & 0.0532 & 0.0304 & 0.0491 & 0.0510 & 0.0562 & 0.0584 & 0.0681 & 0.0742 & 0.0761 & 0.0665 & 0.0806 & 0.0803 & 0.0789 & 0.0794 & 0.0818 & \textbf{0.0873} & 1.5$e^{-6}$ \\
        & NDCG@20 & 0.0264 & 0.0143 & 0.0222 & 0.0245 & 0.0282 & 0.0280 & 0.0336 & 0.0365 & 0.0373 & 0.0327 & 0.0402 & 0.0398 & 0.0391 & 0.0395 & 0.0424 & \textbf{0.0439} & 1.8$e^{-8}$ \\
        \cline{2-19}
        & Recall@40 & 0.0802 & 0.0487 & 0.0692 & 0.0743 & 0.0856 & 0.0891 & 0.1019 & 0.1151 & 0.1175 & 0.1032 & 0.1230 & 0.1226 & 0.1210 & 0.1217 & 0.1226 & \textbf{0.1315} & 3.2$e^{-6}$ \\
        & NDCG@40 & 0.0321 & 0.0187 & 0.0268 & 0.0315 & 0.0355 & 0.0360 & 0.0419 & 0.0466 & 0.0474 & 0.0418 & 0.0505 & 0.0502 & 0.0492 & 0.0497 & 0.0524 & \textbf{0.0548} & 2.7$e^{-7}$ \\
        \hline
        \multirow{4}{*}{BeerAdvocate} & Recall@20 & 0.0996 & 0.0729 & 0.0816 & 0.0930 & 0.1003 & 0.1082 & 0.1033 & 0.1035 & 0.1102 & 0.1048 & 0.1131 & 0.1138 & 0.1156 & 0.1150 & 0.1182 & \textbf{0.1216} & 7.7$e^{-6}$ \\
        & NDCG@20 & 0.0856 & 0.0654 & 0.0650 & 0.0816 & 0.0852 & 0.0901 & 0.0873 & 0.0901 & 0.0943 & 0.0881 & 0.0971 & 0.0959 & 0.0990 & 0.0977 & 0.0981 & \textbf{0.1015} & 4.9$e^{-3}$ \\
        \cline{2-19}
        & Recall@40 & 0.1602 & 0.1203 & 0.1325 & 0.1553 & 0.1650 & 0.1766 & 0.1653 & 0.1662 & 0.1757 & 0.1723 & 0.1819 & 0.1776 & 0.1847 & 0.1799 & 0.1797 & \textbf{0.1867} & 1.3$e^{-2}$ \\
        & NDCG@40 & 0.1016 & 0.0754 & 0.0794 & 0.0980 & 0.1031 & 0.1085 & 0.1032 & 0.1062 & 0.1113 & 0.1068 & 0.1150 & 0.1122 & 0.1176 & 0.1156 & 0.1139 & \textbf{0.1182} & 2.4$e^{-1}$ \\
        \hline
    \end{tabular}
    }    
    \label{tab:exp_comparison}
\end{table*}

\subsection{Experimental Settings}

\subsubsection{\bf Evaluation Datasets.}
We conduct experiments on three datasets collected from online applications, Last.FM, Yelp, and BeerAdvocate. The statistics of these datasets are shown in Table~\ref{tab:dataset}.
\begin{itemize}[leftmargin=*]

    \item \textbf{Last.FM}: This dataset contains social networking, tagging, and music artist listening information collected from a set of users from the Last.fm online music system. \\\vspace{-0.1in}
    
    \item \textbf{Yelp}: This commonly-used dataset contains user ratings on business venues collected from the Yelp platform. It is a valuable resource for studying user preferences and behavior in the context of personalized venue recommendations. \\\vspace{-0.1in}
 %
    

    \item \textbf{BeerAdvocate}: This dataset contains beer reviews from BeerAdvocate. We process it using the 10-core setting by keeping only users and items with at least 10 interactions.

\end{itemize}

\subsubsection{\bf Evaluation Protocols.}

We follow the recent collaborative filtering models~\cite{he2020lightgcn, wu2021self} and split the datasets by 7:2:1 into training, validation, and testing sets. We adopt the all-rank evaluation protocol, where for each test user, the positive items in the test set and all the non-interacted items were tested and ranked together. We employ the commonly-used \textit{Recall@N} and \textit{Normalized Discounted Cumulative Gain (NDCG)@N} as evaluation metrics for recommendation performance evaluation. We set \textit{N} to 20 by default.

\subsubsection{\bf Compared Baseline Methods.}

We evaluate our proposed \model by comparing it with various baselines for comprehensive evaluation. The details of the baselines are as follows.

\begin{itemize}[leftmargin=*]


    \item \textbf{BiasMF}~\cite{koren2009matrix}: It is a matrix factorization method that aims to enhance user-specific preferences for recommendation by incorporating bias vectors for users and items. \\\vspace{-0.1in}


    \item \textbf{NCF}~\cite{he2017neural}: It is a neural network-based method that replaces the dot-product operation in conventional matrix factorization with multi-layer neural networks. This allows the model to capture complex user-item interactions and provide recommendations. For our comparison, we utilize the NeuMF variant of NCF. \\\vspace{-0.1in}

    \item \textbf{AutoR}~\cite{sedhain2015autorec}: It is a method that improves the user/item representations by using a three-layer autoencoder trained under the supervision of an interaction reconstruction task.
    

\end{itemize}






\begin{itemize}[leftmargin=*]


    \item \textbf{GCMC}~\cite{berg2017graph}: This work utilizes graph convolutional networks (GCNs) for interaction matrix completion. \\\vspace{-0.1in}
    
    \item \textbf{PinSage}~\cite{ying2018graph}: It is a graph convolutional-based method that employs random sampling in the graph convolutional framework to enhance the collaborative filtering task. \\\vspace{-0.1in}
    

    \item \textbf{NGCF}~\cite{wang2019neural}: It uses a multi-layer graph convolutional network to propagate information through the user-item interaction graph and learn the latent representations of users and items. \\\vspace{-0.1in}
    
    \item \textbf{STGCN}~\cite{zhang2019star}: It combines graph convolutional encoders with graph autoencoders to enhance the model's robustness against sparse and cold-start samples in collaborative filtering tasks. \\\vspace{-0.1in}
    
    

    \item \textbf{LightGCN}~\cite{he2020lightgcn}: This model leverages the power of neighborhood information in the user-item interaction graph by using a layer-wise propagation scheme that involves only linear transformations and element-wise additions. \\\vspace{-0.1in}
    

    \item \textbf{GCCF}~\cite{chen2020revisiting}: It presents a new approach to collaborative filtering recommender systems by revisiting graph convolutional networks. It removes non-linear activations and introduces a residual network structure that alleviates the over-smoothing problem.
    
\end{itemize}


\begin{itemize}[leftmargin=*]
    \item \textbf{HCCF}~\cite{xia2022hypergraph}: A new self-supervised recommendation framework is proposed in this work, which is able to capture both local and global collaborative relations using a hypergraph neural networks enhanced by cross-view contrastive learning architecture.\\\vspace{-0.1in}
    

    \item \textbf{SHT}~\cite{xia2022self}: It integrates hypergraph neural networks and transformer under a self-supervised learning paradigm for data augmentation to denoise user-item interactions in recommendation.
    
\end{itemize}

\begin{itemize}[leftmargin=*]
    

    \item \textbf{SLRec}~\cite{yao2021self}: It integrates contrastive learning between node features as regularization terms in order to improve the efficacy of current collaborative filtering recommender systems. \\\vspace{-0.1in}
    
    \item \textbf{SGL}~\cite{wu2021self}: The model augments LightGCN with self-supervised contrastive learning by conducting data augmentation through random walk and node/edge dropout to corrupt graph structures. \\\vspace{-0.1in}
    
    \item \textbf{NCL}~\cite{lin2022improving}: This is a neighborhood-enriched contrastive learning approach that enhances graph collaborative filtering by incorporating potential neighbors into contrastive pairs. NCL introduces structural and semantic neighbors of a user or item, developing a structure-contrastive and a prototype-contrastive objective. \\\vspace{-0.1in}


    \item \textbf{DirectAU}~\cite{wang2022towards}: This new approach proposes a new learning objective for collaborative filtering methods that measures the representation quality based on alignment and uniformity on the hypersphere. It directly optimizes these two properties to improve recommendation performance.
        
\end{itemize}

\subsection{Overall Performance Comparison (RQ1)}

The effectiveness of the proposed \model is validated through an overall performance evaluation on three datasets, comparing it with various baselines. To ensure statistical significance, the authors retrained \model and the best-performing baseline five times and computed p-values. The results are presented in Table~\ref{tab:exp_comparison}.

\begin{itemize}[leftmargin=*]
    

    \item The evaluation results indicate that \model outperforms the baselines under both top-\textit{20} and top-\textit{40} settings, and the t-tests validate the significance of the observed performance improvements. The superior performance of \model can be attributed to the effectiveness of the proposed contrastive learning frameworks for data augmentation over user-item interactions. The use of adaptive view generators ensures that informative and diverse contrastive views are generated. This, in turn, leads to more effective learning of user and item embeddings, resulting in better recommendations. Overall, these findings demonstrate the effectiveness of the proposed contrastive learning approach for collorative filtering and highlight the importance of designing effective data augmentation techniques for this task. \\\vspace{-0.1in}
    
    %

    \item 
    The evaluation results demonstrate that self-supervised learning improves existing CF frameworks, such as SLRec, SGL, and NCL. This improvement can be attributed to incorporating an augmented learning task, which provides beneficial regularization based on the input data. For example, SLRec and SGL use stochastic data augmentation to generate multiple views, while NCL incorporates potential neighbors into contrastive pairs. However, these methods may lose useful signals that reflect important user-item interaction patterns. In contrast, \model has two main advantages. First, it does not rely on random data augmentation to generate contrastive views, instead using two adaptive view generators to create reasonable views that retain useful information. The generative view captures the key patterns of the original data, while the denoising generator filters out noise signals that may interfere with the contrastive learning process.

    Second, \model addresses the problem of model collapse in contrastive learning by creating contrastive views from different aspects with two different generators. The generative and denoising views capture different aspects of the input data, ensuring that the learned representations are diverse and informative. The superior performance of \model compared to the baseline self-supervised approaches validates the effectiveness of this new self-supervised learning paradigm for CF.

\end{itemize}

\begin{table}[t]
    \centering
    \caption{Ablation study on key components of \model.}
    \vspace{-0.1in}
    \footnotesize
    \setlength{\tabcolsep}{0.8mm}
    \resizebox{\linewidth}{!}{
    \begin{tabular}{c|c|c c|c c|c c}
        \hline
        \multirow{2}{*}{Category} & Data & \multicolumn{2}{|c|}{Last.FM} & \multicolumn{2}{|c|}{Yelp} & \multicolumn{2}{|c}{BeerAdvocate} \\
        \cline{2-8}
        & Variants & Recall & NDCG & Recall & NDCG & Recall & NDCG \\
        \hline
        \hline
        \multirow{2}{*}{Adaptive} 
         & w/o Task & 0.2562 & 0.1868 & 0.0849 & 0.0425 & 0.1212 & 0.1010 \\
         & Gen+Gen & 0.2494 & 0.1819 & 0.0853 & 0.0429 & 0.1187 & 0.0992 \\
         \hline
         \multirow{1}{*}{Random} & EdgeD & 0.2476 & 0.1794 & 0.0852 & 0.0424 & 0.1163 & 0.0964 \\
         \hline
         \multicolumn{2}{c|}{\model} & 0.2603 & 0.1911 & 0.0873 & 0.0439 & 0.1216 & 0.1015 \\
        \hline
    \end{tabular}
    }    
    \label{tab:exp_ablation}
    \vspace{-0.1in}
\end{table}

\subsection{Model Ablation Test (RQ2)}


We conducted extensive experiments to validate the effectiveness of the proposed methods by removing three applied techniques in \model individually: the adaptive view generators, the task-aware optimization for view generators, and the denoising view generator.

To evaluate the efficacy of the proposed generative and denoising generators for view generation, we compare them to existing random augmentation method. Specifically, an ablated version of \model is trained using the random edge drop augmentation ($\emph{EdgeD}$). Additionally, we replace the denoising view generator with an identical VGAE-based generator ($\emph{Gen+Gen}$), to study the importance of denoising in the view generation process.
Furthermore, we replace the task-aware optimization with the original reconstruction objective ($\emph{w/o Task}$), to investigate the necessity of introducing task-relevant information into model training.
The variants are re-trained and tested on the three datasets. The results are presented in Table~\ref{tab:exp_ablation}, from which we draw the following major conclusions:

\begin{itemize}[leftmargin=*]
    \item \textbf{Advantage of adaptive view generators}.
    The results presented in Table~\ref{tab:exp_ablation} demonstrate that using the random-permutation-based contrastive view generator ($\emph{EdgeD}$) leads to a significant decay in performance compared to the proposed \model approach. This suggests that random augmentation methods may not be sufficient for generating informative contrastive views in CF. In contrast, the adaptive learning ability of the generative view based on VGAE and the denoising ability of the explicit denoising network in \model are critical for achieving superior performance. The generative view preserves the key patterns of the original data by modeling the graph-based user-item interaction structures, while the denoising network filters out noise signals that may interfere with the contrastive learning process. \\\vspace{-0.1in}

    \item \textbf{Benefit of denoising view generator}. We conduct additional tests on a modified version of our model to further study the effectiveness of our designed adaptive view generators. Specifically, we remove the denoising view generator (referred to as the $\emph{Gen+Gen}$ variant). The results show that, while the VGAE-based view provide adaptive data augmentations that benefit contrastive learning, it is not enough to eliminate the inherent data noise. Our \model addresses this issue by incorporating the denoising view into the contrastive learning process, resulting in significant performance improvements.\\\vspace{-0.1in}
    
    \item \textbf{Effectiveness of the task-aware optimization}. The results show that the \textit{w/o Task} variant performs worse than the proposed \model on all three datasets. This suggests that using a general-purpose auto-encoding loss and denoising loss for contrastive view generator training may not be sufficient for achieving optimal performance in CF. Instead, introducing BPR loss for task-aware view generator training leads to better performance. This highlights the importance of incorporating task-aware information to guide the training of view generators, which can help to capture more relevant user-item interaction patterns and improve the quality of the generated contrastive views. \\\vspace{-0.1in}

\end{itemize}

\subsection{Model Robustness Test (RQ3)}
In this section, our experiments show that our proposed approach, \model, exhibits superior robustness against data noise, and is effective in handling sparse user-item interaction data.

\subsubsection{\bf Performance \textit{w.r.t.} Data Noise Degree.}

\begin{figure}[h]
    \centering
    \subfigure[Lasat.FM data]{
    \includegraphics[width=0.49 \linewidth]{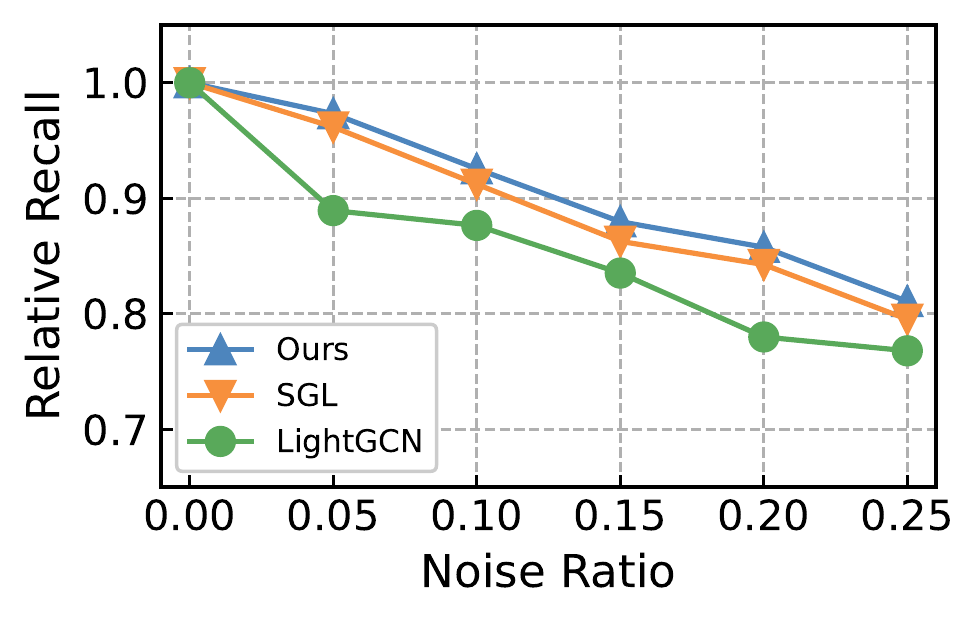}
    \includegraphics[width=0.49 \linewidth]{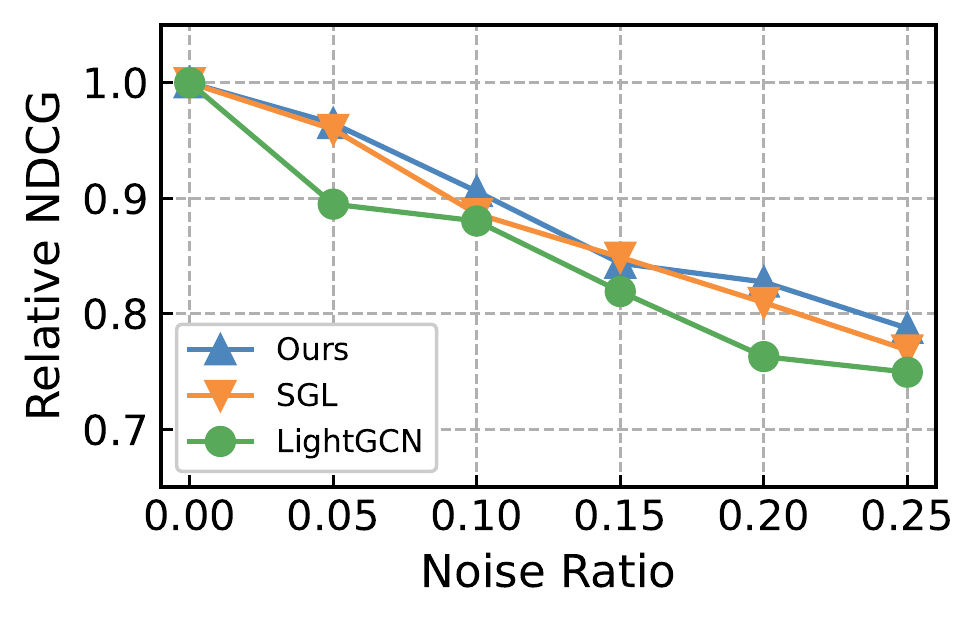}}
    \subfigure[Yelp data]{
    \includegraphics[width=0.49 \linewidth]{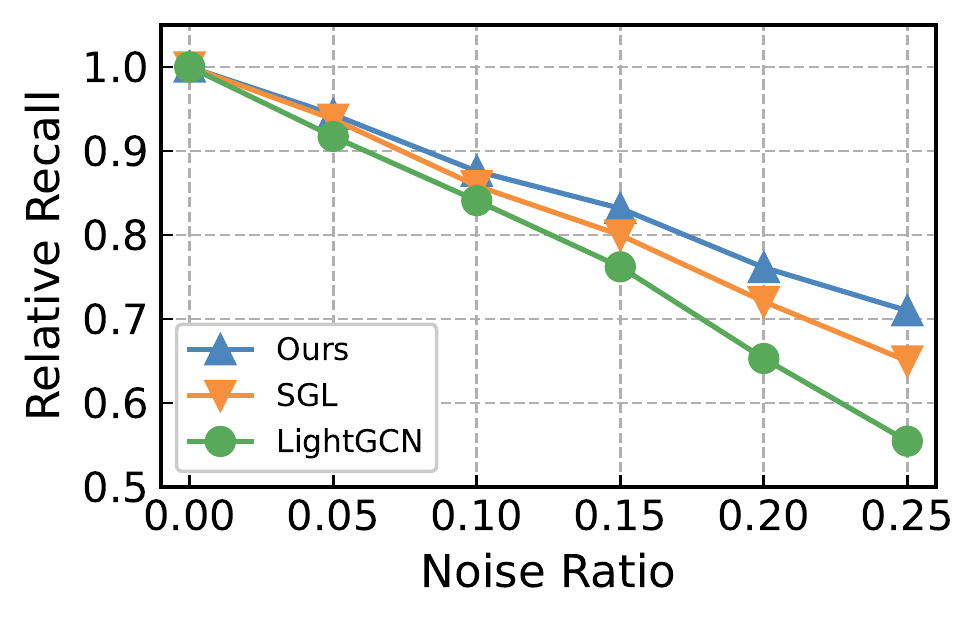}
    \includegraphics[width=0.49 \linewidth]{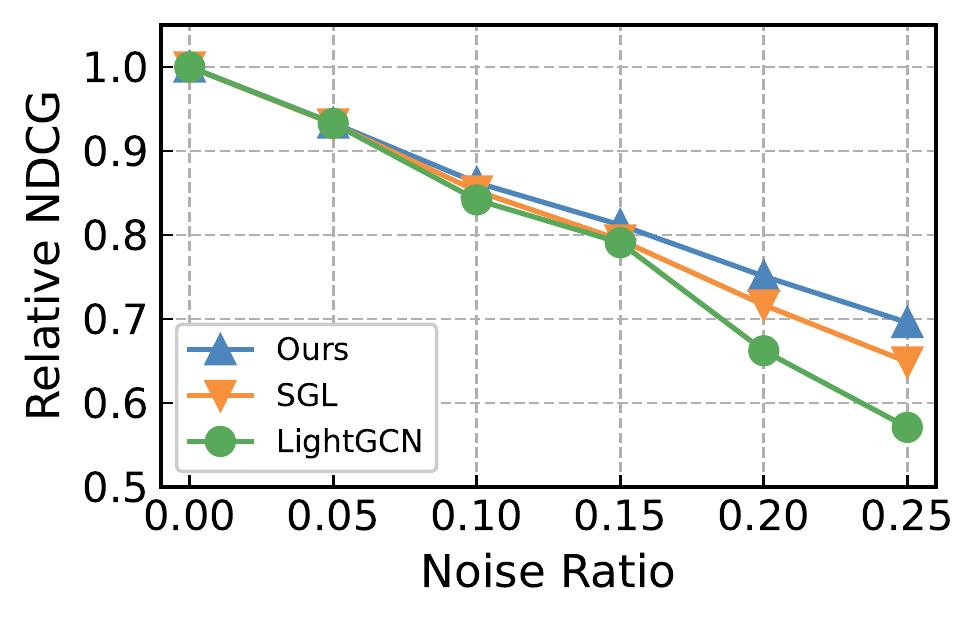}}
    \subfigure[BeerAdvocate data]{
    \includegraphics[width=0.49 \linewidth]{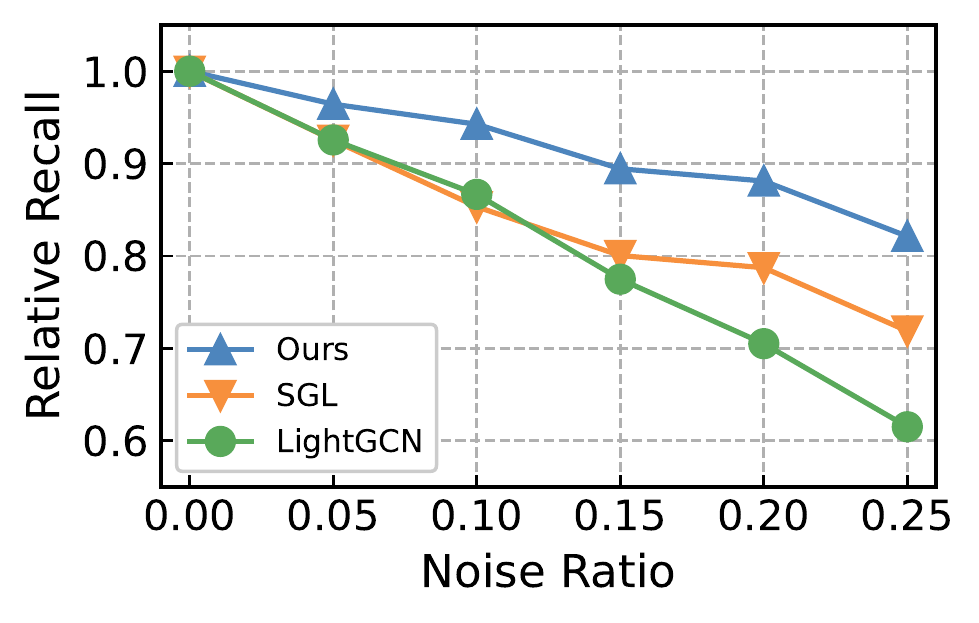}
    \includegraphics[width=0.49 \linewidth]{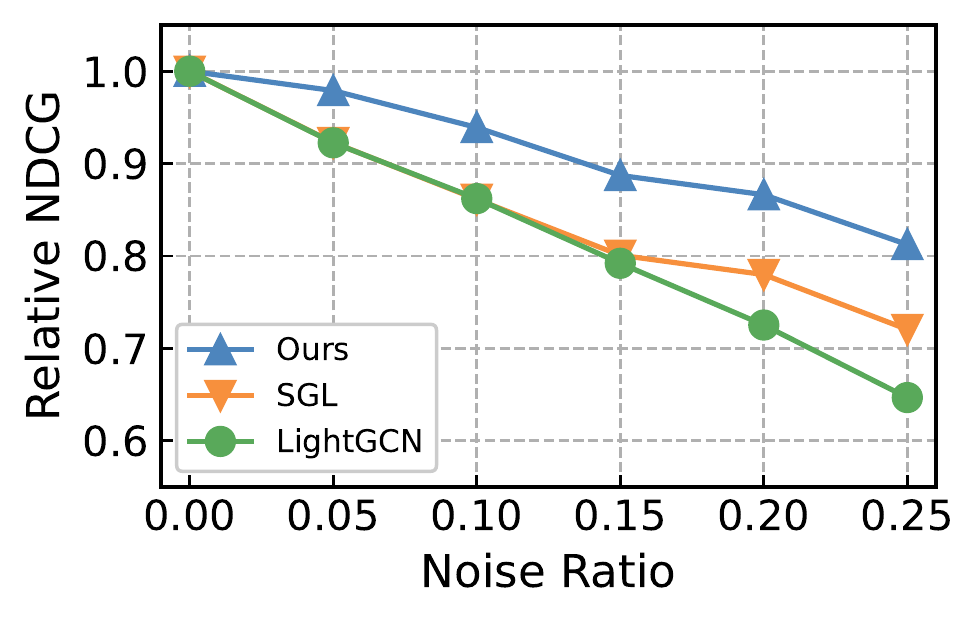}}
    \vspace{-0.15in}
    \caption{Relative performance degradation \textit{w.r.t.} noise ratio. We introduce varying levels of noise by replacing 5\%, 10\%, 15\%, 20\%, and 25\% of the interaction edges with fake edges.}
    \label{fig:figure_exp_noise}
    \vspace{-0.1in}
\end{figure}


We investigate the robustness of our approach, \model, against data noise in recommendation systems. To evaluate the impact of noise on our model's performance, we randomly replace different percentages of real edges with fake edges and retrain the model using the corrupted graphs as input. Concretely, we replace 5\%, 10\%, 15\%, 20\%, and 25\% of the interaction edges with fake edges in our experiments. We compare \model's performance with two other models, LightGCN and SGL. To better understand the effect of noise on performance degradation, we evaluate the relative performance compared to the performance on the original data, and present the results in Fig.~\ref{fig:figure_exp_noise}. Our observations indicate that \model exhibits smaller performance degradation in most cases compared to the baselines.

We attribute this observation to two reasons: First, the self-supervised learning task employed by \model distills information from two adaptive contrastive views to refine the graph embeddings. This observation is supported by the stronger robustness of the self-supervised method SGL compared to LightGCN. Second, both view generators used in our approach are capable of generating a contrastive view with less noise and more task-related information. Additionally, we find that the relative performance degradation on the Yelp dataset is more apparent compared to the other two datasets. This finding is because noisy data has a larger influence on the performance of models on sparse datasets like Yelp, which is the sparest dataset in our experiments. Overall, our results suggest that \model is a robust and effective model for recommendation systems, even in the presence of data noise.

\subsubsection{\bf Performance \textit{w.r.t.} Data Sparsity.}

\begin{figure}[h]
    \centering
    \subfigure[Performance \textit{w.r.t.} user interaction numbers]{
    \includegraphics[width=0.5 \linewidth]{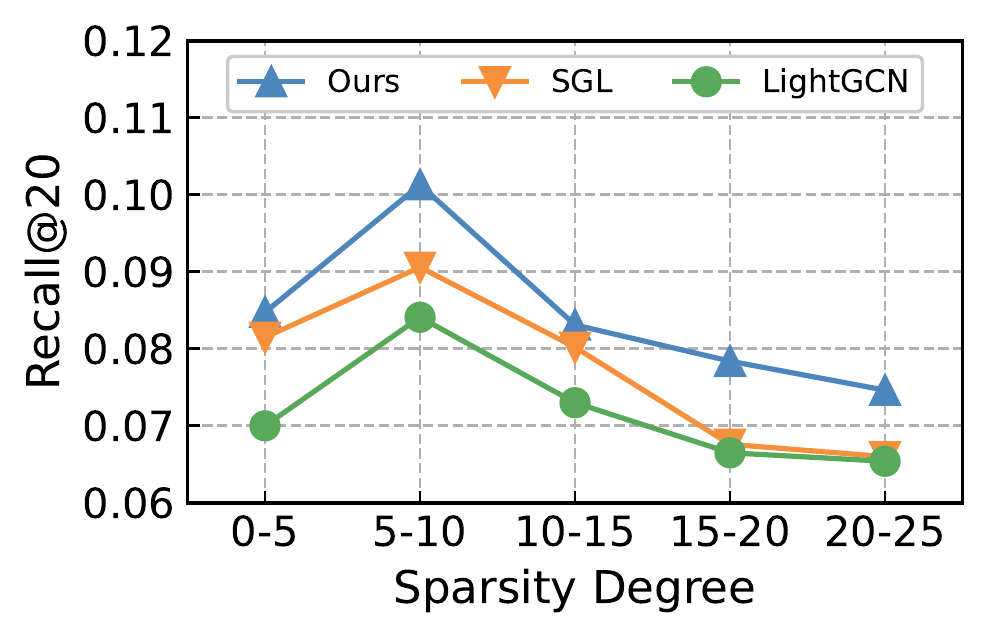}
    \includegraphics[width=0.5 \linewidth]{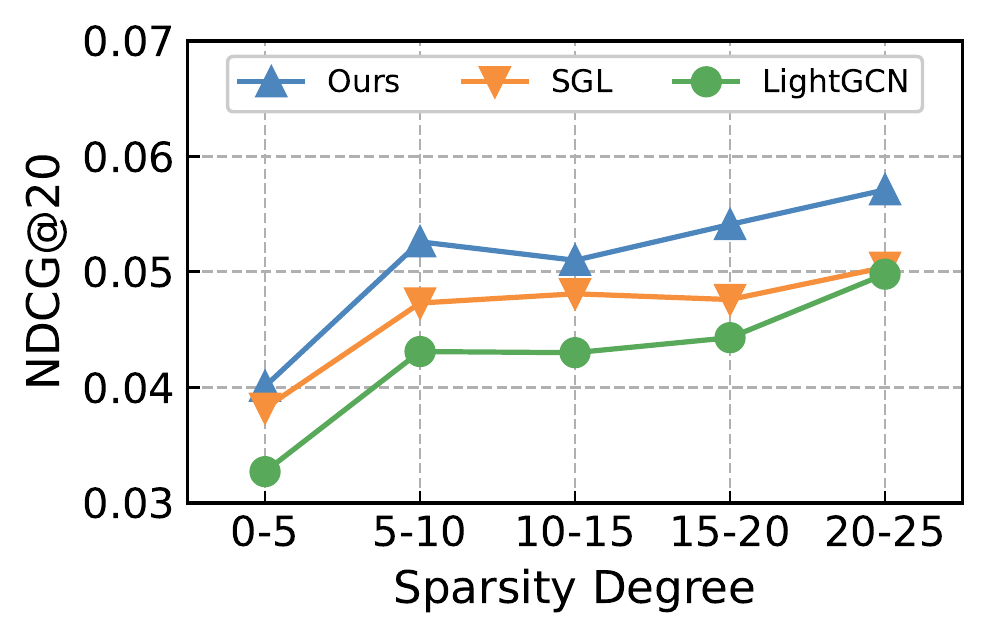}}
    \subfigure[Performance \textit{w.r.t.} item interaction numbers]{
    \includegraphics[width=0.5 \linewidth]{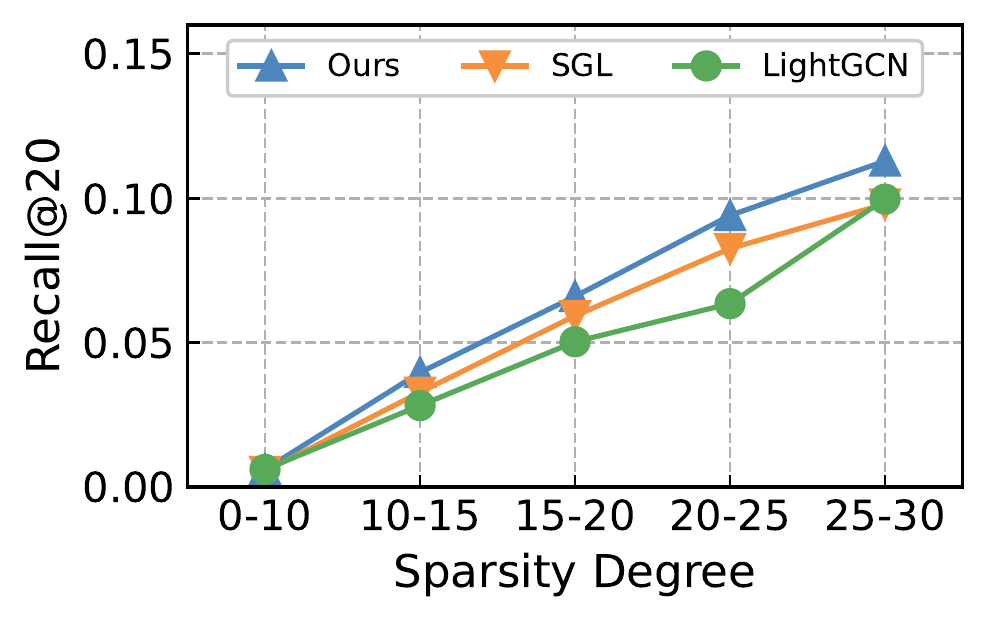}
    \includegraphics[width=0.5 \linewidth]{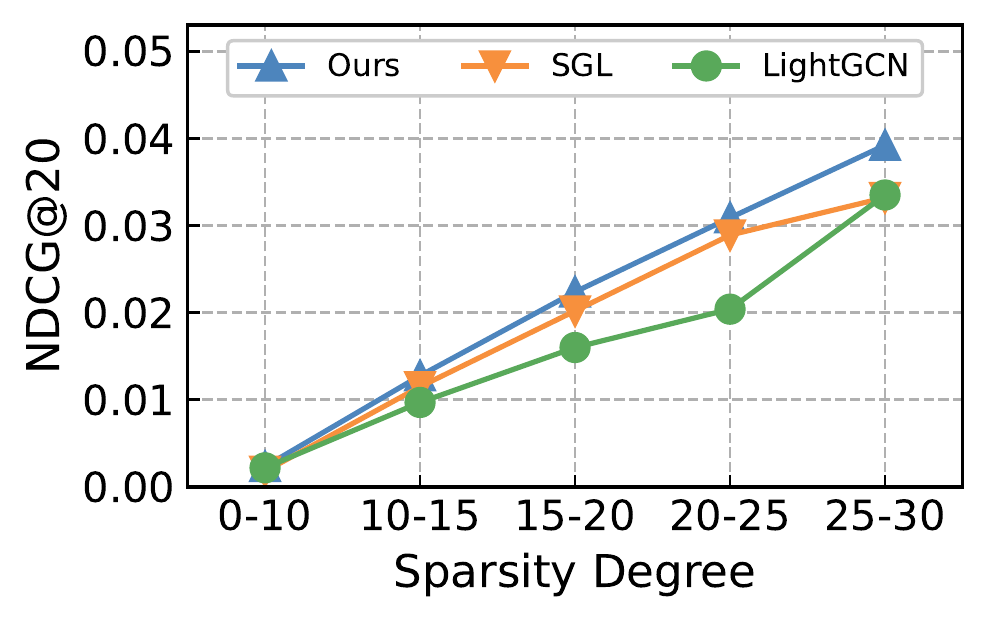}}
    \vspace{-0.15in}
    \caption{Performance \textit{w.r.t} different sparsity degrees of interaction data for users and items, respectively, on Yelp dataset. We divide users and items into several groups based on the number of interactions they had in the dataset.}
    \label{fig:figure_exp_sparsity}
    \vspace{-0.1in}
\end{figure}



We also investigate the influence of data sparsity on model performance from both user and item sides. We compare our proposed \model with LightGCN and SGL in this experiment. Multiple user and item groups are constructed based on their number of interactions in the training set, with the first group in the user-side experiments containing users interacting with 0-10 items and the first group in the item-side experiments containing items interacting with 0-5 users. 

Fig.~\ref{fig:figure_exp_sparsity} illustrates the recommendation accuracy for our \model and the two compared methods. Our findings highlight the following: First, \model exhibits consistently superior performance on datasets with different sparsity degrees, indicating its robustness in handling sparse data for both users and items. We attribute this advantage to our adaptive contrastive view pair, which provides high-quality self-supervised signals that mitigate the negative effects of data sparsity. Second, the sparsity of item interaction vectors has a more significant influence on model performance across all the methods. Overall, our experiments demonstrate the effectiveness of \model in handling sparse user-item interaction data. 

\subsection{Hyperparameter Analysis (RQ4)}

\begin{figure}[t]
    \centering
    \subfigure[Recall@20]{
    \includegraphics[width=0.48 \linewidth]{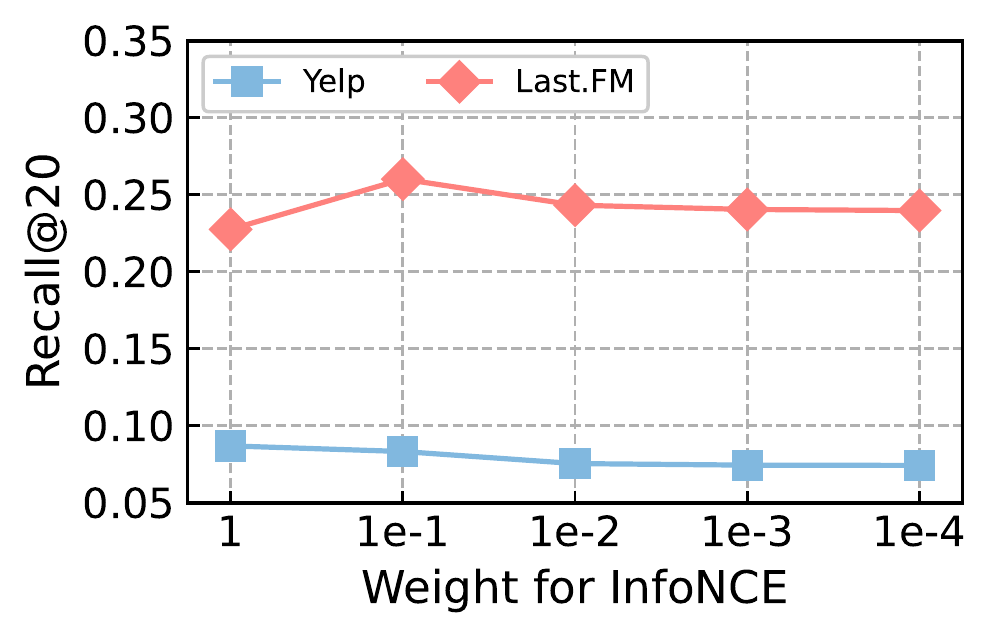}}
    \subfigure[NDCG@20]{
    \includegraphics[width=0.48 \linewidth]{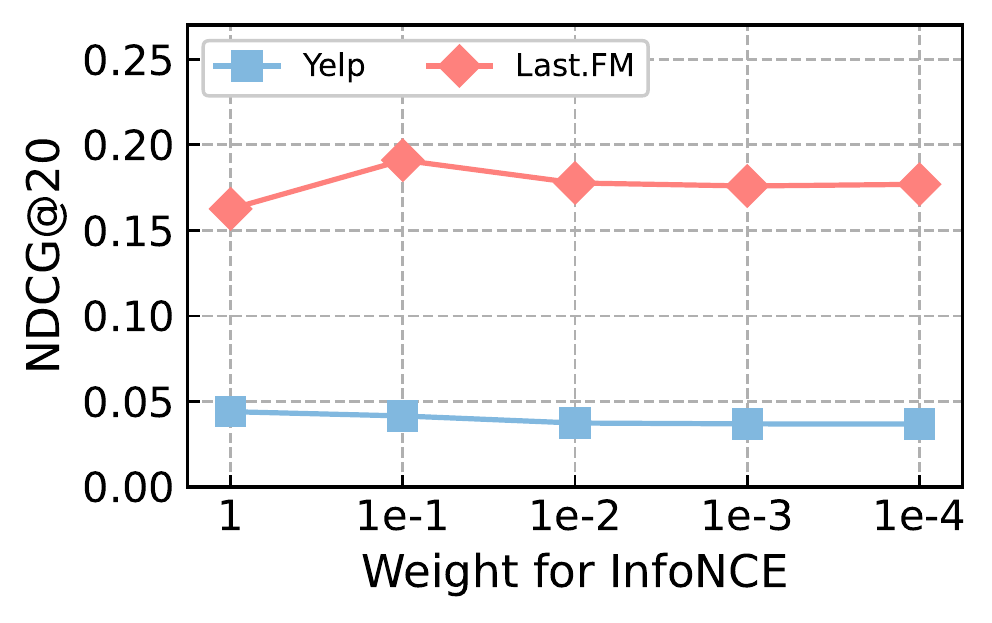}}
    \vspace{-0.1in}
    \caption{Hyperparameter Analysis on Last.FM and Yelp.}
    \label{fig:figure_param}
    \vspace{-0.1in}
\end{figure}

In this section, the authors investigate the sensitivity of their proposed model to the key hyperparameter $\lambda_1$ for InfoNCE loss, which controls the strength of contrastive learning. Specifically, the weight $\lambda_1$ is searched in the range of ($1$, $1e^{-1}$, $1e^{-2}$, $1e^{-3}$, $1e^{-4}$) to explore its impact on the model's performance. The results are presented in Figure~\ref{fig:figure_param}, which shows the model's performance on the Last.FM and Yelp datasets with different values of $\lambda_1$. It is observed that the best performance is achieved with $\lambda_1 = 1e-1$ and $\lambda_1 = 1$. This suggests that a large value of $\lambda_1$ may overly emphasize the contrastive optimization loss.

\begin{figure*}[t]
    \centering
    \vspace{-0.1in}
    \subfigure[Main View]{
    \label{fig:figure_vis_ours_main}
    \includegraphics[width=0.15 \linewidth]{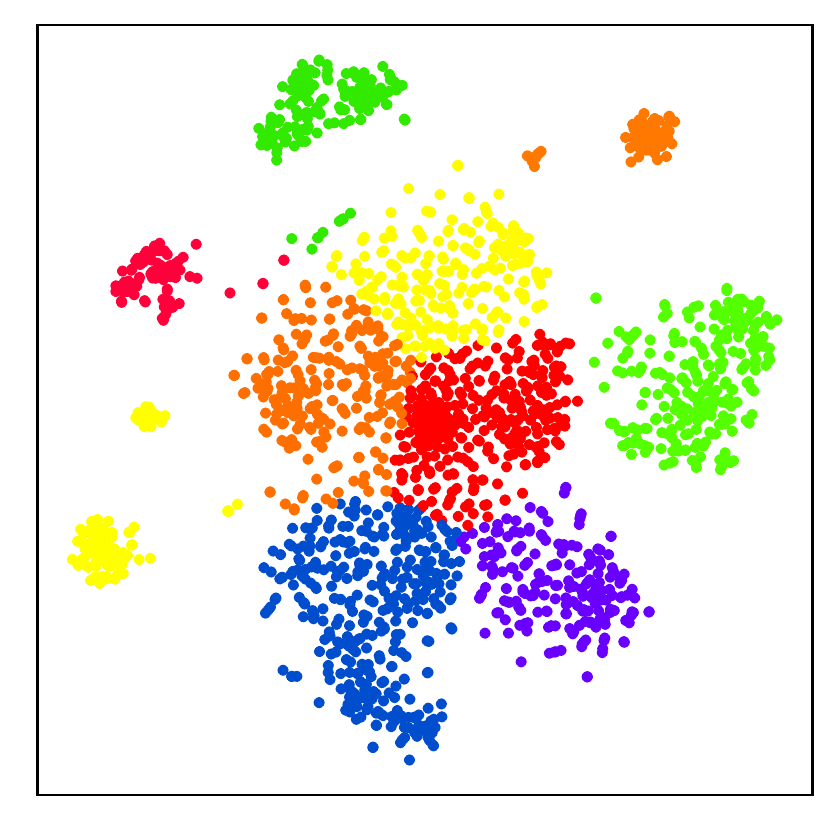}}
    \subfigure[CL View 1]{
    \label{fig:figure_vis_ours_1}
    \includegraphics[width=0.15 \linewidth]{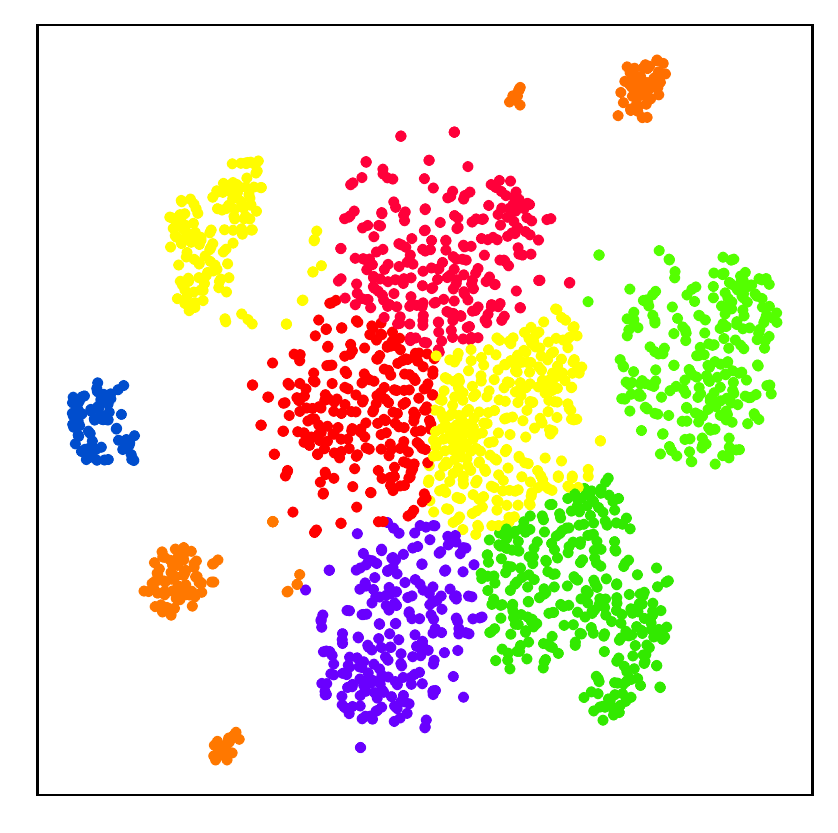}}
    \subfigure[CL View 2]{
    \label{fig:figure_vis_ours_2}
    \includegraphics[width=0.15 \linewidth]{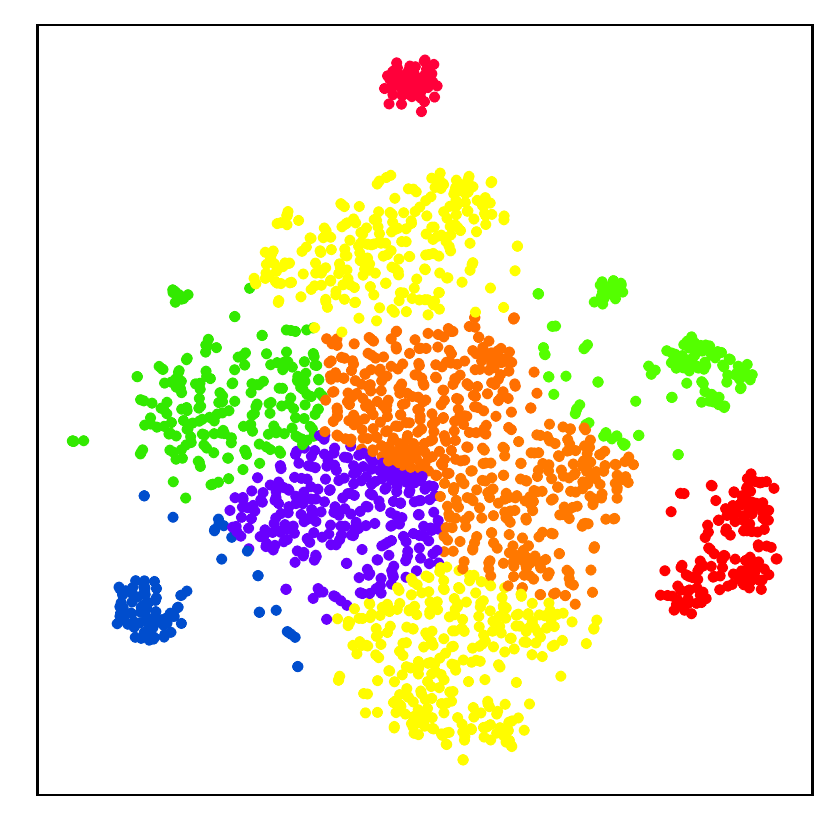}}
    \subfigure[Noisy Main View]{
    \label{fig:figure_vis_ours_main_noise}
    \includegraphics[width=0.15 \linewidth]{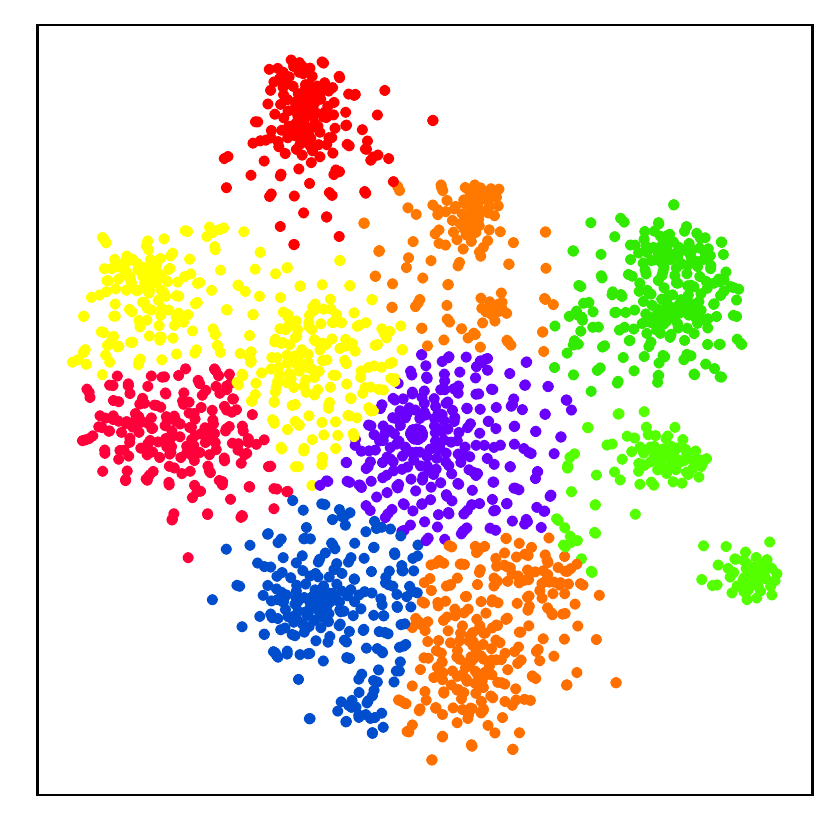}}
    \subfigure[Noisy CL View 1]{
    \label{fig:figure_vis_ours_1_noise}
    \includegraphics[width=0.15 \linewidth]{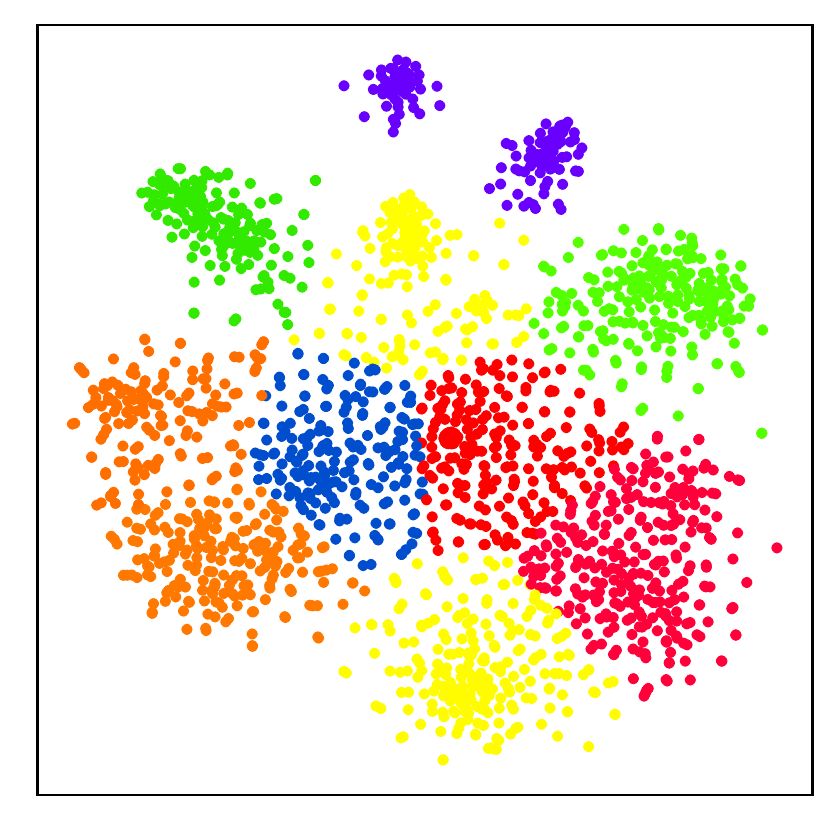}}
    \subfigure[Noisy CL View 2]{
    \label{fig:figure_vis_ours_2_noise}
    \includegraphics[width=0.15 \linewidth]{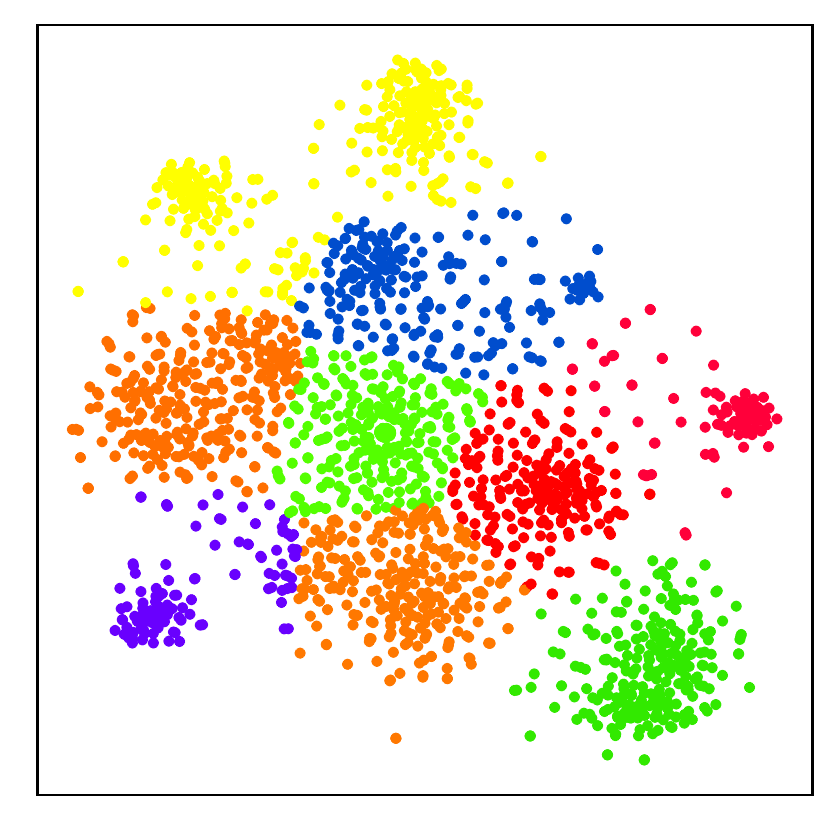}}
    \vspace{-0.1in}
    \caption{View embedding visualization for \model.}
    \label{fig:figure_vis_ours}
\end{figure*}

\begin{figure*}[t]
    \centering
    \vspace{-0.1in}
    \subfigure[Main View]{
    \label{fig:figure_vis_sgl_main}
    \includegraphics[width=0.15 \linewidth]{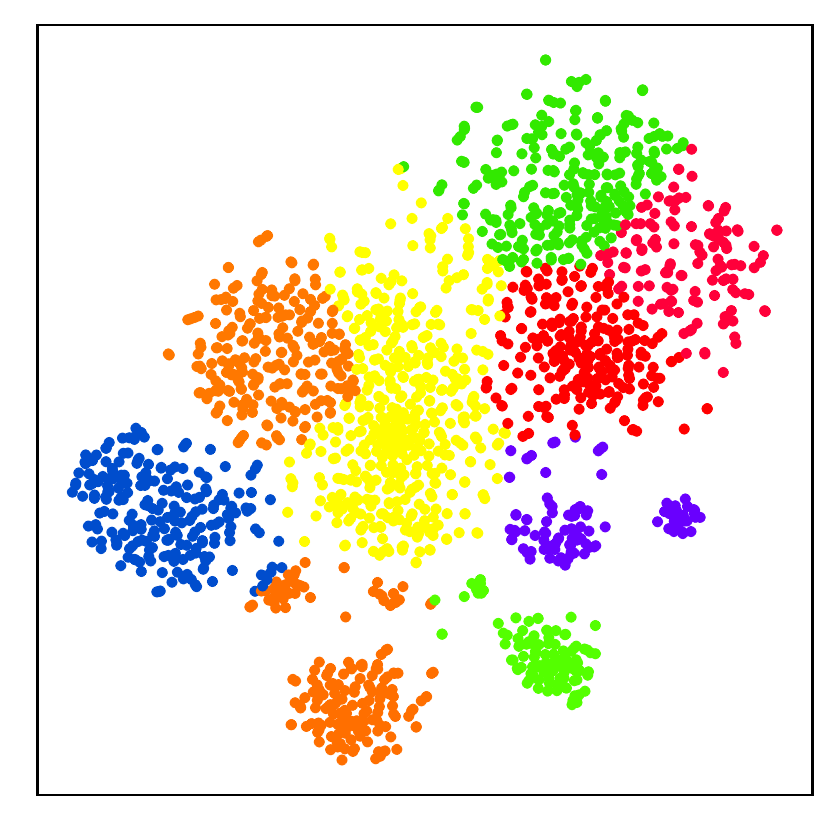}}
    \subfigure[CL View 1]{
    \label{fig:figure_vis_sgl_1}
    \includegraphics[width=0.15 \linewidth]{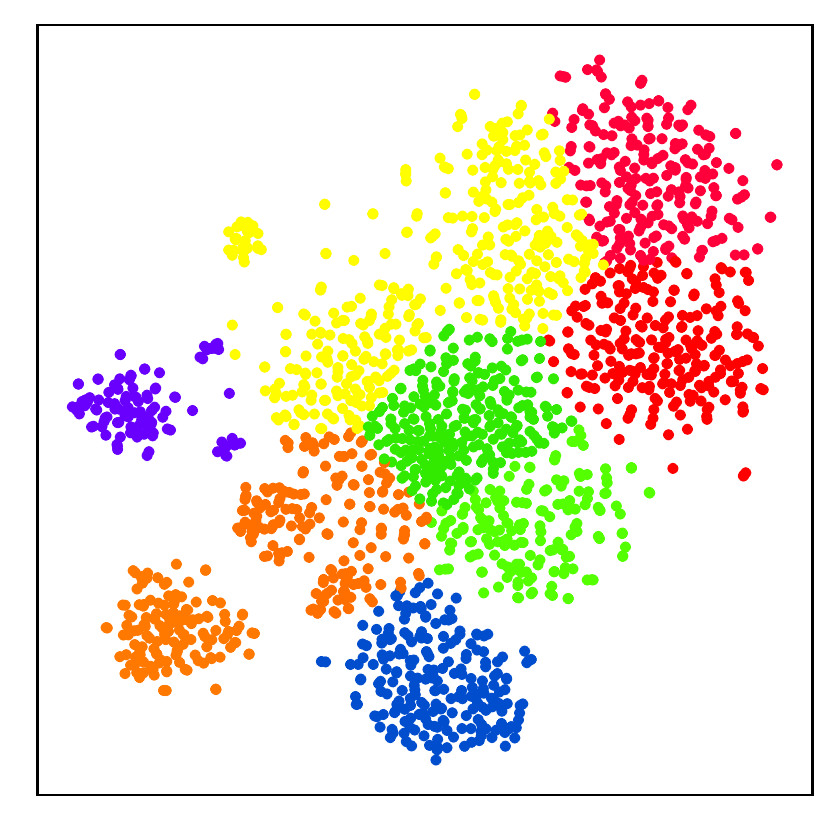}}
    \subfigure[CL View 2]{
    \label{fig:figure_vis_sgl_2}
    \includegraphics[width=0.15 \linewidth]{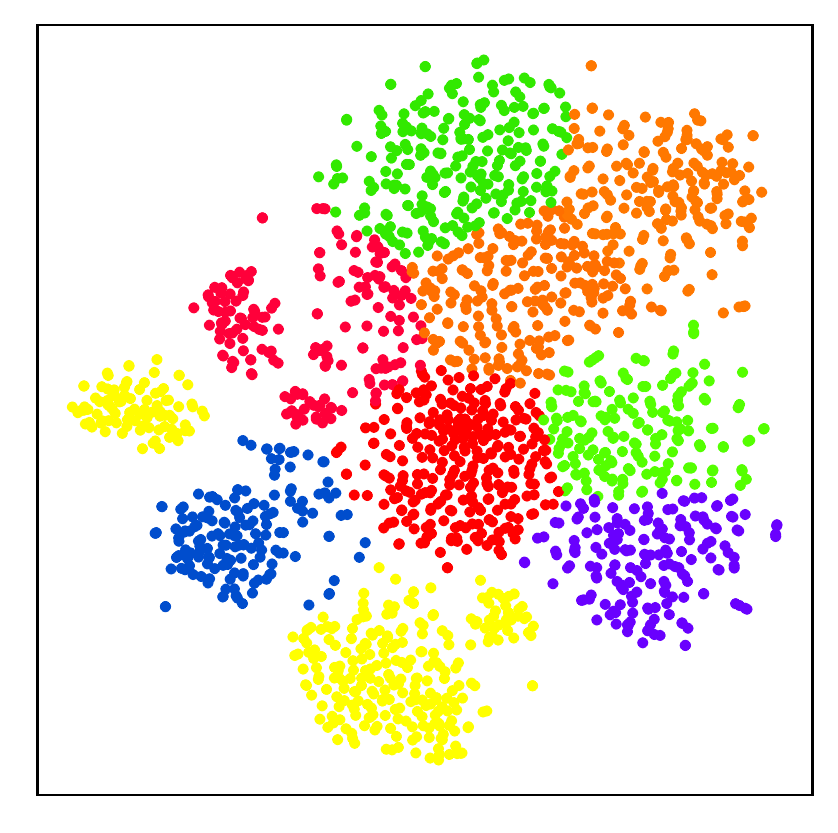}}
    \subfigure[Noisy Main View]{
    \label{fig:figure_vis_sgl_main_noise}
    \includegraphics[width=0.15 \linewidth]{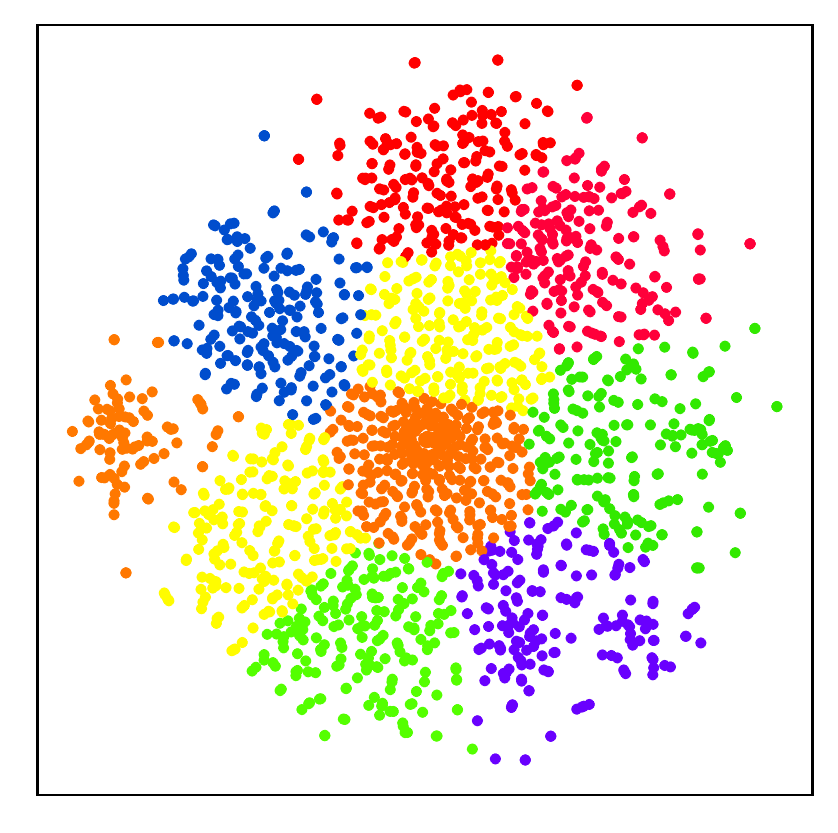}}
    \subfigure[Noisy CL View 1]{
    \label{fig:figure_vis_sgl_1_noise}
    \includegraphics[width=0.15 \linewidth]{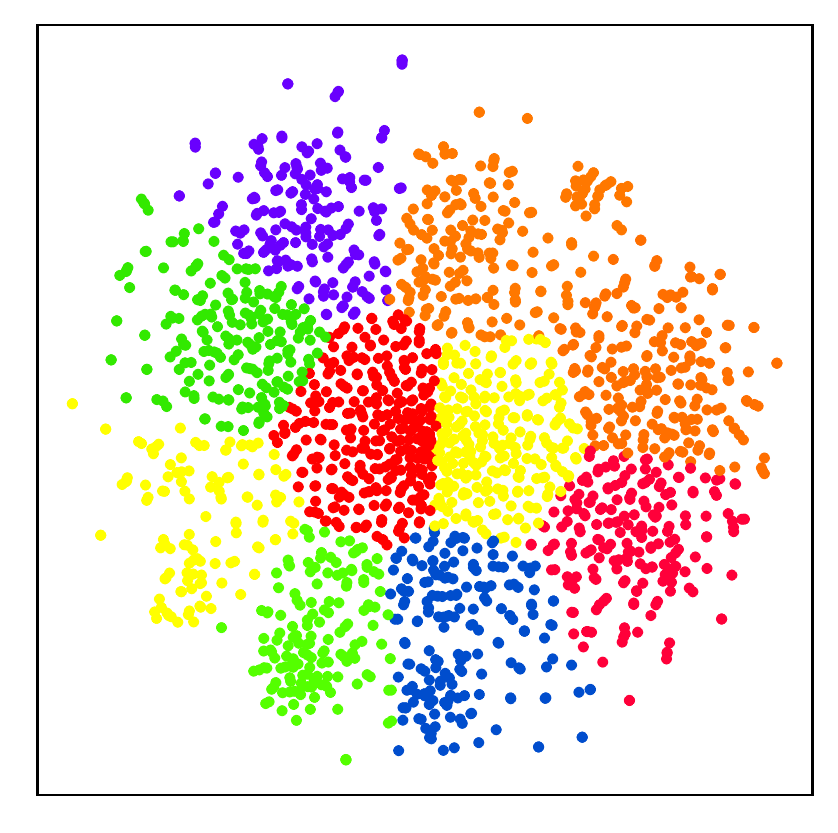}}
    \subfigure[Noisy CL View 2]{
    \label{fig:figure_vis_sgl_2_noise}
    \includegraphics[width=0.15 \linewidth]{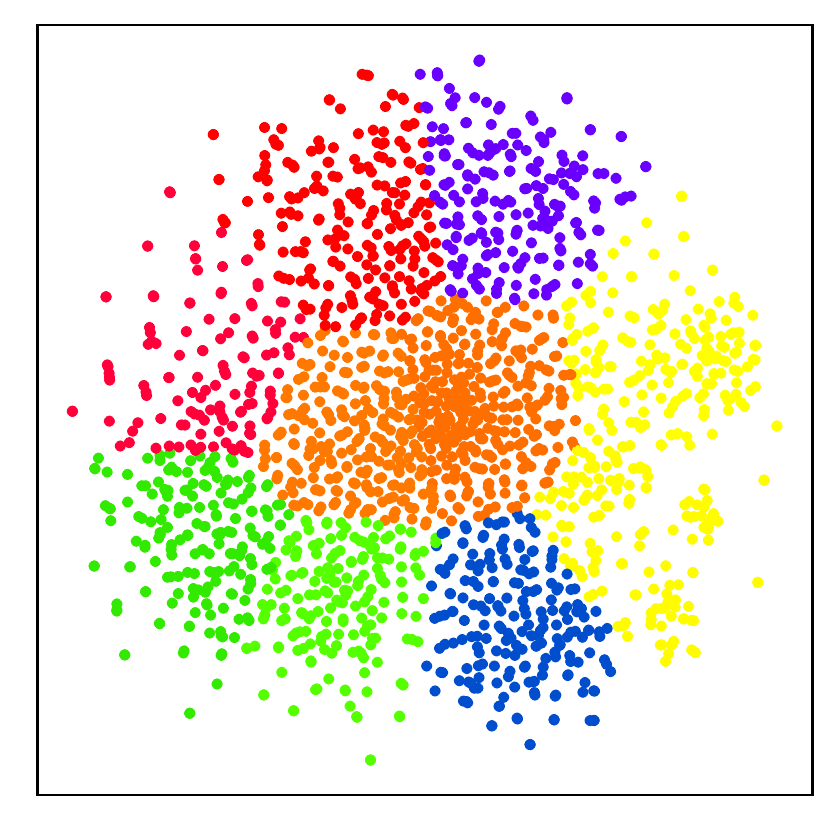}}
    \vspace{-0.1in}
    \caption{View embedding visualization for SGL.}
    \label{fig:figure_vis_sgl}
    \vspace{-0.1in}
\end{figure*}

\subsection{Embedding Visualisation Analysis}

In this section, we conduct an embedding visualization analysis with the representations encoded from our proposed approach, AdaGCL, and the baseline SGL to gain insight into the benefits of our model. As previously mentioned, SGL uses random data augmentation methods to create contrastive views, which can result in poor performance when dealing with noisy data. The added noises may unintentionally cause damage to contrastive views. Furthermore, SGL employs the same data augmentation methods on both contrastive views, leading to the issue of model collapse since the two views can easily have a similar distribution.

To validate the effectiveness of our method in addressing these limitations, we visualize the embeddings of the two contrastive views given by AdaGCL and SGL. We randomly sample 2,000 nodes from the Yelp dataset and map their embeddings in the three views (i.e., one main view and two contrastive views) to the 2-D space with t-SNE~\cite{van2008visualizing}. We employ the KMeans algorithm to cluster the nodes based on their compressed 2-D embeddings and color them with different colors. To highlight the impact of noisy data on SGL and AdaGCL, we also visualize the embeddings of polluted data, where 25\% of the edges are replaced with fake edges. The visualization results are shown in Fig.\ref{fig:figure_vis_ours} and Fig.\ref{fig:figure_vis_sgl}, respectively.
Note that in Fig.\ref{fig:figure_vis_ours}, View 1 and View 2 are generated by the graph generative model and the graph denoising model, respectively.

\subsubsection{\bf Effectiveness of Adaptive View Generators.}

As shown in Fig.\ref{fig:figure_vis_sgl_main}, SGL learns a large cloud of evenly-distanced embeddings with a few clear community structures to capture the collaborative relations among nodes. This is because random edge dropping tends to generate contrastive views with uniform distributions, as shown in Fig.\ref{fig:figure_vis_sgl_1} and Fig.~\ref{fig:figure_vis_sgl_2}. Furthermore, SGL's two contrastive views show more similar distributions compared to our method. In contrast, our \model is based on two adaptive view generators that can generate more informative and diverse views of the data. By adaptively adjusting the views, our method is able to capture more complex and nuanced structures of the graph, resulting in more distinct embeddings with better clustering effects.

Furthermore, our \model demonstrates better robustness when dealing with noisy data compared to SGL. The visualization results of the three views of SGL (\textit{i.e.}, Fig.\ref{fig:figure_vis_sgl_main_noise}, Fig.\ref{fig:figure_vis_sgl_1_noise}, and Fig.~\ref{fig:figure_vis_sgl_2_noise}) show severe over-uniform distributions. When dealing with noisy data, SGL can produce embeddings with uniform distributions, which can result in a loss of unique collaborative patterns in the embeddings and negatively impact the performance of the method. In contrast, according to the visual results of the three views of our \model (\textit{i.e.}, Fig.\ref{fig:figure_vis_ours_main_noise}, Fig.\ref{fig:figure_vis_ours_1_noise}, and Fig.~\ref{fig:figure_vis_ours_2_noise}), our method is more robust to noisy data. This is because our method adaptively adjusts the views to capture the most informative and discriminative aspects of the data, and is therefore less affected by noise in the input.

\subsubsection{\bf Effectiveness of the Graph Denoising Model.}

To improve the robustness of our \model to noise and avoid the issue of model collapse, we design a graph denoising module as the second view generator. As shown in Fig.\ref{fig:figure_vis_ours_2} and Fig.\ref{fig:figure_vis_ours_2_noise}, the contrastive view created by the graph denoising component is less affected by noise compared to the other view pair (\textit{i.e.}, Fig.\ref{fig:figure_vis_ours_1} and Fig.\ref{fig:figure_vis_ours_1_noise}). This is because the denoised graph contains more informative and discriminative signals about the graph structure corresponding to complex user-item interaction patterns, making it more robust to noise. Moreover, our graph denoising model creates better denoised views compared to the contrastive views in SGL (\textit{i.e.}, Fig.\ref{fig:figure_vis_sgl_1_noise} and Fig.\ref{fig:figure_vis_sgl_2_noise}), validating its effectiveness in graph denoising. By incorporating the denoised graph as augmented view, the captured more informative signals resulting in more robust user representations.

\section{CONCLUSION}


In this work, we propose a novel approach to improving contrastive recommender systems through the use of adaptive view generators. Specifically, we introduce a new recommendation framework, \model, which utilizes a graph generative model and a graph denoising model to create contrastive views, allowing for more effective user-item interaction modeling with self-augmented supervision signals. Our framework demonstrates improved robustness against noise perturbation, thereby enhancing the overall performance of graph-based recommender systems. Through extensive experimentation on multiple datasets, we have shown that our proposed \model, outperforms several competitive baselines, providing validation for its superiority in contrastive recommenders. 

Moving forward, an important area of research would be to extend our framework to explore casual factors for contrastive self-supervised learning signals in recommender systems. This involves leveraging causal inference techniques~\cite{wang2022causal} to improve the interpretability of the self-supervised learning signals used in contrastive learning. By accounting for the underlying causal relationships between user behaviors, we can design more effective and informative self-supervised learning objectives that better capture the nuances of user-item interactions. 
Additionally, we may investigate the transferability of our model by exploring transfer learning techniques, such as domain adaptation and multi-task learning.
\clearpage




\clearpage
\bibliographystyle{ACM-Reference-Format}
\balance
\bibliography{main.bbl}


\begin{thebibliography}{40}


\ifx \showCODEN    \undefined \def \showCODEN     #1{\unskip}     \fi
\ifx \showDOI      \undefined \def \showDOI       #1{#1}\fi
\ifx \showISBNx    \undefined \def \showISBNx     #1{\unskip}     \fi
\ifx \showISBNxiii \undefined \def \showISBNxiii  #1{\unskip}     \fi
\ifx \showISSN     \undefined \def \showISSN      #1{\unskip}     \fi
\ifx \showLCCN     \undefined \def \showLCCN      #1{\unskip}     \fi
\ifx \shownote     \undefined \def \shownote      #1{#1}          \fi
\ifx \showarticletitle \undefined \def \showarticletitle #1{#1}   \fi
\ifx \showURL      \undefined \def \showURL       {\relax}        \fi
\providecommand\bibfield[2]{#2}
\providecommand\bibinfo[2]{#2}
\providecommand\natexlab[1]{#1}
\providecommand\showeprint[2][]{arXiv:#2}

\bibitem[Berg et~al\mbox{.}(2017)]%
        {berg2017graph}
\bibfield{author}{\bibinfo{person}{Rianne van~den Berg},
  \bibinfo{person}{Thomas~N Kipf}, {and} \bibinfo{person}{Max Welling}.}
  \bibinfo{year}{2017}\natexlab{}.
\newblock \showarticletitle{Graph convolutional matrix completion}.
\newblock \bibinfo{journal}{\emph{arXiv preprint arXiv:1706.02263}}
  (\bibinfo{year}{2017}).
\newblock


\bibitem[Cao et~al\mbox{.}(2022)]%
        {cao2022contrastive}
\bibfield{author}{\bibinfo{person}{Jiangxia Cao}, \bibinfo{person}{Xin Cong},
  \bibinfo{person}{Jiawei Sheng}, \bibinfo{person}{Tingwen Liu}, {and}
  \bibinfo{person}{Bin Wang}.} \bibinfo{year}{2022}\natexlab{}.
\newblock \showarticletitle{Contrastive Cross-Domain Sequential
  Recommendation}. In \bibinfo{booktitle}{\emph{International Conference on
  Information \& Knowledge Management (CIKM)}}. \bibinfo{pages}{138--147}.
\newblock


\bibitem[Chen et~al\mbox{.}(2020)]%
        {chen2020revisiting}
\bibfield{author}{\bibinfo{person}{Lei Chen}, \bibinfo{person}{Le Wu},
  \bibinfo{person}{Richang Hong}, \bibinfo{person}{Kun Zhang}, {and}
  \bibinfo{person}{Meng Wang}.} \bibinfo{year}{2020}\natexlab{}.
\newblock \showarticletitle{Revisiting graph based collaborative filtering: A
  linear residual graph convolutional network approach}. In
  \bibinfo{booktitle}{\emph{AAAI Conference on Artificial Intelligence
  (AAAI)}}, Vol.~\bibinfo{volume}{34}. \bibinfo{pages}{27--34}.
\newblock


\bibitem[He et~al\mbox{.}(2020)]%
        {he2020lightgcn}
\bibfield{author}{\bibinfo{person}{Xiangnan He}, \bibinfo{person}{Kuan Deng},
  \bibinfo{person}{Xiang Wang}, \bibinfo{person}{Yan Li},
  \bibinfo{person}{Yongdong Zhang}, {and} \bibinfo{person}{Meng Wang}.}
  \bibinfo{year}{2020}\natexlab{}.
\newblock \showarticletitle{Lightgcn: Simplifying and powering graph
  convolution network for recommendation}. In
  \bibinfo{booktitle}{\emph{International Conference on Research and
  Development in Information Retrieval (SIGIR)}}. \bibinfo{pages}{639--648}.
\newblock


\bibitem[He et~al\mbox{.}(2017)]%
        {he2017neural}
\bibfield{author}{\bibinfo{person}{Xiangnan He}, \bibinfo{person}{Lizi Liao},
  \bibinfo{person}{Hanwang Zhang}, \bibinfo{person}{Liqiang Nie},
  \bibinfo{person}{Xia Hu}, {and} \bibinfo{person}{Tat-Seng Chua}.}
  \bibinfo{year}{2017}\natexlab{}.
\newblock \showarticletitle{Neural collaborative filtering}. In
  \bibinfo{booktitle}{\emph{The Web Conference (WWW)}}.
  \bibinfo{pages}{173--182}.
\newblock


\bibitem[Hsu and Li(2021)]%
        {hsu2021retagnn}
\bibfield{author}{\bibinfo{person}{Cheng Hsu} {and} \bibinfo{person}{Cheng-Te
  Li}.} \bibinfo{year}{2021}\natexlab{}.
\newblock \showarticletitle{Retagnn: Relational temporal attentive graph neural
  networks for holistic sequential recommendation}. In
  \bibinfo{booktitle}{\emph{The Web Conference (WWW)}}.
  \bibinfo{pages}{2968--2979}.
\newblock


\bibitem[Hwang et~al\mbox{.}(2020)]%
        {hwang2020self}
\bibfield{author}{\bibinfo{person}{Dasol Hwang}, \bibinfo{person}{Jinyoung
  Park}, \bibinfo{person}{Sunyoung Kwon}, \bibinfo{person}{KyungMin Kim},
  \bibinfo{person}{Jung-Woo Ha}, {and} \bibinfo{person}{Hyunwoo~J Kim}.}
  \bibinfo{year}{2020}\natexlab{}.
\newblock \showarticletitle{Self-supervised auxiliary learning with meta-paths
  for heterogeneous graphs}. In \bibinfo{booktitle}{\emph{Advances in Neural
  Information Processing Systems (NeurIPS)}}. \bibinfo{pages}{10294--10305}.
\newblock


\bibitem[Jamali and Ester(2010)]%
        {jamali2010matrix}
\bibfield{author}{\bibinfo{person}{Mohsen Jamali} {and} \bibinfo{person}{Martin
  Ester}.} \bibinfo{year}{2010}\natexlab{}.
\newblock \showarticletitle{A matrix factorization technique with trust
  propagation for recommendation in social networks}. In
  \bibinfo{booktitle}{\emph{ACM Conference on Recommender Systems (Recsys)}}.
  \bibinfo{pages}{135--142}.
\newblock


\bibitem[Kipf and Welling(2016)]%
        {kipf2016variational}
\bibfield{author}{\bibinfo{person}{Thomas~N Kipf} {and} \bibinfo{person}{Max
  Welling}.} \bibinfo{year}{2016}\natexlab{}.
\newblock \showarticletitle{Variational graph auto-encoders}.
\newblock \bibinfo{journal}{\emph{arXiv preprint arXiv:1611.07308}}
  (\bibinfo{year}{2016}).
\newblock


\bibitem[Koren et~al\mbox{.}(2009)]%
        {koren2009matrix}
\bibfield{author}{\bibinfo{person}{Yehuda Koren}, \bibinfo{person}{Robert
  Bell}, {and} \bibinfo{person}{Chris Volinsky}.}
  \bibinfo{year}{2009}\natexlab{}.
\newblock \showarticletitle{Matrix factorization techniques for recommender
  systems}.
\newblock \bibinfo{journal}{\emph{Computer}} \bibinfo{volume}{42},
  \bibinfo{number}{8} (\bibinfo{year}{2009}), \bibinfo{pages}{30--37}.
\newblock


\bibitem[Li et~al\mbox{.}(2023)]%
        {li2023graph}
\bibfield{author}{\bibinfo{person}{Chaoliu Li}, \bibinfo{person}{Lianghao Xia},
  \bibinfo{person}{Xubin Ren}, \bibinfo{person}{Yaowen Ye},
  \bibinfo{person}{Yong Xu}, {and} \bibinfo{person}{Chao Huang}.}
  \bibinfo{year}{2023}\natexlab{}.
\newblock \showarticletitle{Graph Transformer for Recommendation}.
\newblock \bibinfo{journal}{\emph{arXiv preprint arXiv:2306.02330}}
  (\bibinfo{year}{2023}).
\newblock


\bibitem[Lin et~al\mbox{.}(2022)]%
        {lin2022improving}
\bibfield{author}{\bibinfo{person}{Zihan Lin}, \bibinfo{person}{Changxin Tian},
  \bibinfo{person}{Yupeng Hou}, {and} \bibinfo{person}{Wayne~Xin Zhao}.}
  \bibinfo{year}{2022}\natexlab{}.
\newblock \showarticletitle{Improving Graph Collaborative Filtering with
  Neighborhood-enriched Contrastive Learning}. In \bibinfo{booktitle}{\emph{The
  Web Conference (WWW)}}. \bibinfo{pages}{2320--2329}.
\newblock


\bibitem[Ren et~al\mbox{.}(2023)]%
        {ren2023disentangled}
\bibfield{author}{\bibinfo{person}{Xubin Ren}, \bibinfo{person}{Lianghao Xia},
  \bibinfo{person}{Jiashu Zhao}, \bibinfo{person}{Dawei Yin}, {and}
  \bibinfo{person}{Chao Huang}.} \bibinfo{year}{2023}\natexlab{}.
\newblock \showarticletitle{Disentangled Contrastive Collaborative Filtering}.
\newblock \bibinfo{journal}{\emph{arXiv preprint arXiv:2305.02759}}
  (\bibinfo{year}{2023}).
\newblock


\bibitem[Sedhain et~al\mbox{.}(2015)]%
        {sedhain2015autorec}
\bibfield{author}{\bibinfo{person}{Suvash Sedhain},
  \bibinfo{person}{Aditya~Krishna Menon}, \bibinfo{person}{Scott Sanner}, {and}
  \bibinfo{person}{Lexing Xie}.} \bibinfo{year}{2015}\natexlab{}.
\newblock \showarticletitle{Autorec: Autoencoders meet collaborative
  filtering}. In \bibinfo{booktitle}{\emph{The Web Conference (WWW)}}.
  \bibinfo{pages}{111--112}.
\newblock


\bibitem[Tao et~al\mbox{.}(2022)]%
        {tao2022self}
\bibfield{author}{\bibinfo{person}{Zhulin Tao}, \bibinfo{person}{Xiaohao Liu},
  \bibinfo{person}{Yewei Xia}, \bibinfo{person}{Xiang Wang},
  \bibinfo{person}{Lifang Yang}, \bibinfo{person}{Xianglin Huang}, {and}
  \bibinfo{person}{Tat-Seng Chua}.} \bibinfo{year}{2022}\natexlab{}.
\newblock \showarticletitle{Self-supervised learning for multimedia
  recommendation}.
\newblock \bibinfo{journal}{\emph{Transactions on Multimedia (TMM)}}
  (\bibinfo{year}{2022}).
\newblock


\bibitem[Van~der Maaten and Hinton(2008)]%
        {van2008visualizing}
\bibfield{author}{\bibinfo{person}{Laurens Van~der Maaten} {and}
  \bibinfo{person}{Geoffrey Hinton}.} \bibinfo{year}{2008}\natexlab{}.
\newblock \showarticletitle{Visualizing data using t-SNE.}
\newblock \bibinfo{journal}{\emph{Journal of machine learning research}}
  \bibinfo{volume}{9}, \bibinfo{number}{11} (\bibinfo{year}{2008}).
\newblock


\bibitem[Wang et~al\mbox{.}(2022b)]%
        {wang2022towards}
\bibfield{author}{\bibinfo{person}{Chenyang Wang}, \bibinfo{person}{Yuanqing
  Yu}, \bibinfo{person}{Weizhi Ma}, \bibinfo{person}{Min Zhang},
  \bibinfo{person}{Chong Chen}, \bibinfo{person}{Yiqun Liu}, {and}
  \bibinfo{person}{Shaoping Ma}.} \bibinfo{year}{2022}\natexlab{b}.
\newblock \showarticletitle{Towards Representation Alignment and Uniformity in
  Collaborative Filtering}. In \bibinfo{booktitle}{\emph{International
  Conference on Knowledge Discovery and Data Mining (KDD)}}.
  \bibinfo{pages}{1816--1825}.
\newblock


\bibitem[Wang et~al\mbox{.}(2019b)]%
        {wang2019learning}
\bibfield{author}{\bibinfo{person}{Hongwei Wang}, \bibinfo{person}{Jialin
  Wang}, \bibinfo{person}{Jia Wang}, \bibinfo{person}{Miao Zhao},
  \bibinfo{person}{Weinan Zhang}, \bibinfo{person}{Fuzheng Zhang},
  \bibinfo{person}{Wenjie Li}, \bibinfo{person}{Xing Xie}, {and}
  \bibinfo{person}{Minyi Guo}.} \bibinfo{year}{2019}\natexlab{b}.
\newblock \showarticletitle{Learning graph representation with generative
  adversarial nets}.
\newblock \bibinfo{journal}{\emph{Transactions on Knowledge and Data
  Engineering (TKDE)}} \bibinfo{volume}{33}, \bibinfo{number}{8}
  (\bibinfo{year}{2019}), \bibinfo{pages}{3090--3103}.
\newblock


\bibitem[Wang et~al\mbox{.}(2020b)]%
        {wang2020time}
\bibfield{author}{\bibinfo{person}{Jianling Wang}, \bibinfo{person}{Raphael
  Louca}, \bibinfo{person}{Diane Hu}, \bibinfo{person}{Caitlin Cellier},
  \bibinfo{person}{James Caverlee}, {and} \bibinfo{person}{Liangjie Hong}.}
  \bibinfo{year}{2020}\natexlab{b}.
\newblock \showarticletitle{Time to Shop for Valentine's Day: Shopping
  Occasions and Sequential Recommendation in E-commerce}. In
  \bibinfo{booktitle}{\emph{International Conference on Web Search and Data
  Mining (WSDM)}}. \bibinfo{pages}{645--653}.
\newblock


\bibitem[Wang et~al\mbox{.}(2022a)]%
        {wang2022causal}
\bibfield{author}{\bibinfo{person}{Wenjie Wang}, \bibinfo{person}{Xinyu Lin},
  \bibinfo{person}{Fuli Feng}, \bibinfo{person}{Xiangnan He},
  \bibinfo{person}{Min Lin}, {and} \bibinfo{person}{Tat-Seng Chua}.}
  \bibinfo{year}{2022}\natexlab{a}.
\newblock \showarticletitle{Causal representation learning for
  out-of-distribution recommendation}. In \bibinfo{booktitle}{\emph{The Web
  Conference (WWW)}}. \bibinfo{pages}{3562--3571}.
\newblock


\bibitem[Wang et~al\mbox{.}(2019a)]%
        {wang2019neural}
\bibfield{author}{\bibinfo{person}{Xiang Wang}, \bibinfo{person}{Xiangnan He},
  \bibinfo{person}{Meng Wang}, \bibinfo{person}{Fuli Feng}, {and}
  \bibinfo{person}{Tat-Seng Chua}.} \bibinfo{year}{2019}\natexlab{a}.
\newblock \showarticletitle{Neural graph collaborative filtering}. In
  \bibinfo{booktitle}{\emph{International Conference on Research and
  Development in Information Retrieval (SIGIR)}}. \bibinfo{pages}{165--174}.
\newblock


\bibitem[Wang et~al\mbox{.}(2020a)]%
        {wang2020disentangled}
\bibfield{author}{\bibinfo{person}{Xiang Wang}, \bibinfo{person}{Hongye Jin},
  \bibinfo{person}{An Zhang}, \bibinfo{person}{Xiangnan He},
  \bibinfo{person}{Tong Xu}, {and} \bibinfo{person}{Tat-Seng Chua}.}
  \bibinfo{year}{2020}\natexlab{a}.
\newblock \showarticletitle{Disentangled graph collaborative filtering}. In
  \bibinfo{booktitle}{\emph{International Conference on Research and
  Development in Information Retrieval (SIGIR)}}. \bibinfo{pages}{1001--1010}.
\newblock


\bibitem[Wang et~al\mbox{.}(2020c)]%
        {wang2020global}
\bibfield{author}{\bibinfo{person}{Ziyang Wang}, \bibinfo{person}{Wei Wei},
  \bibinfo{person}{Gao Cong}, \bibinfo{person}{Xiao-Li Li},
  \bibinfo{person}{Xian-Ling Mao}, {and} \bibinfo{person}{Minghui Qiu}.}
  \bibinfo{year}{2020}\natexlab{c}.
\newblock \showarticletitle{Global context enhanced graph neural networks for
  session-based recommendation}. In \bibinfo{booktitle}{\emph{International
  Conference on Research and Development in Information Retrieval (SIGIR)}}.
  \bibinfo{pages}{169--178}.
\newblock


\bibitem[Wang et~al\mbox{.}(2022c)]%
        {wang2022learning}
\bibfield{author}{\bibinfo{person}{Zhaobo Wang}, \bibinfo{person}{Yanmin Zhu},
  \bibinfo{person}{Haobing Liu}, {and} \bibinfo{person}{Chunyang Wang}.}
  \bibinfo{year}{2022}\natexlab{c}.
\newblock \showarticletitle{Learning graph-based disentangled representations
  for next POI recommendation}. In \bibinfo{booktitle}{\emph{International
  Conference on Research and Development in Information Retrieval (SIGIR)}}.
  \bibinfo{pages}{1154--1163}.
\newblock


\bibitem[Wei et~al\mbox{.}(2023)]%
        {2023mmssl}
\bibfield{author}{\bibinfo{person}{Wei Wei}, \bibinfo{person}{Chao Huang},
  \bibinfo{person}{Lianghao Xia}, {and} \bibinfo{person}{Chuxu Zhang}.}
  \bibinfo{year}{2023}\natexlab{}.
\newblock \showarticletitle{Multi-Modal Self-Supervised Learning for
  Recommendation}. In \bibinfo{booktitle}{\emph{The Web Conference (WWW)}}.
  \bibinfo{pages}{790--800}.
\newblock


\bibitem[Wu et~al\mbox{.}(2021)]%
        {wu2021self}
\bibfield{author}{\bibinfo{person}{Jiancan Wu}, \bibinfo{person}{Xiang Wang},
  \bibinfo{person}{Fuli Feng}, \bibinfo{person}{Xiangnan He},
  \bibinfo{person}{Liang Chen}, \bibinfo{person}{Jianxun Lian}, {and}
  \bibinfo{person}{Xing Xie}.} \bibinfo{year}{2021}\natexlab{}.
\newblock \showarticletitle{Self-supervised graph learning for recommendation}.
  In \bibinfo{booktitle}{\emph{International Conference on Research and
  Development in Information Retrieval (SIGIR)}}. \bibinfo{pages}{726--735}.
\newblock


\bibitem[Wu et~al\mbox{.}(2018)]%
        {wu2018turning}
\bibfield{author}{\bibinfo{person}{Liang Wu}, \bibinfo{person}{Diane Hu},
  \bibinfo{person}{Liangjie Hong}, {and} \bibinfo{person}{Huan Liu}.}
  \bibinfo{year}{2018}\natexlab{}.
\newblock \showarticletitle{Turning clicks into purchases: Revenue optimization
  for product search in e-commerce}. In \bibinfo{booktitle}{\emph{International
  Conference on Research \& Development in Information Retrieval (SIGIR)}}.
  \bibinfo{pages}{365--374}.
\newblock


\bibitem[Wu et~al\mbox{.}(2022)]%
        {wu2022graph}
\bibfield{author}{\bibinfo{person}{Shiwen Wu}, \bibinfo{person}{Fei Sun},
  \bibinfo{person}{Wentao Zhang}, \bibinfo{person}{Xu Xie}, {and}
  \bibinfo{person}{Bin Cui}.} \bibinfo{year}{2022}\natexlab{}.
\newblock \showarticletitle{Graph neural networks in recommender systems: a
  survey}.
\newblock \bibinfo{journal}{\emph{Comput. Surveys}} \bibinfo{volume}{55},
  \bibinfo{number}{5} (\bibinfo{year}{2022}), \bibinfo{pages}{1--37}.
\newblock


\bibitem[Xia et~al\mbox{.}(2023)]%
        {xia2023graph}
\bibfield{author}{\bibinfo{person}{Lianghao Xia}, \bibinfo{person}{Chao Huang},
  \bibinfo{person}{Jiao Shi}, {and} \bibinfo{person}{Yong Xu}.}
  \bibinfo{year}{2023}\natexlab{}.
\newblock \showarticletitle{Graph-less collaborative filtering}. In
  \bibinfo{booktitle}{\emph{The Web Conference (WWW)}}.
  \bibinfo{pages}{17--27}.
\newblock


\bibitem[Xia et~al\mbox{.}(2022b)]%
        {xia2022hypergraph}
\bibfield{author}{\bibinfo{person}{Lianghao Xia}, \bibinfo{person}{Chao Huang},
  \bibinfo{person}{Yong Xu}, \bibinfo{person}{Jiashu Zhao},
  \bibinfo{person}{Dawei Yin}, {and} \bibinfo{person}{Jimmy Huang}.}
  \bibinfo{year}{2022}\natexlab{b}.
\newblock \showarticletitle{Hypergraph contrastive collaborative filtering}. In
  \bibinfo{booktitle}{\emph{International Conference on Research and
  Development in Information Retrieval (SIGIR)}}. \bibinfo{pages}{70--79}.
\newblock


\bibitem[Xia et~al\mbox{.}(2022a)]%
        {xia2022self}
\bibfield{author}{\bibinfo{person}{Lianghao Xia}, \bibinfo{person}{Chao Huang},
  {and} \bibinfo{person}{Chuxu Zhang}.} \bibinfo{year}{2022}\natexlab{a}.
\newblock \showarticletitle{Self-supervised hypergraph transformer for
  recommender systems}. In \bibinfo{booktitle}{\emph{International Conference
  on Knowledge Discovery and Data Mining (KDD)}}. \bibinfo{pages}{2100--2109}.
\newblock


\bibitem[Yao et~al\mbox{.}(2021)]%
        {yao2021self}
\bibfield{author}{\bibinfo{person}{Tiansheng Yao}, \bibinfo{person}{Xinyang
  Yi}, \bibinfo{person}{Derek~Zhiyuan Cheng}, \bibinfo{person}{Felix Yu},
  \bibinfo{person}{Ting Chen}, \bibinfo{person}{Aditya Menon},
  \bibinfo{person}{Lichan Hong}, \bibinfo{person}{Ed~H Chi},
  \bibinfo{person}{Steve Tjoa}, \bibinfo{person}{Jieqi Kang}, {et~al\mbox{.}}}
  \bibinfo{year}{2021}\natexlab{}.
\newblock \showarticletitle{Self-supervised learning for large-scale item
  recommendations}. In \bibinfo{booktitle}{\emph{International Conference on
  Information \& Knowledge Management (CIKM)}}. \bibinfo{pages}{4321--4330}.
\newblock


\bibitem[Ying et~al\mbox{.}(2018)]%
        {ying2018graph}
\bibfield{author}{\bibinfo{person}{Rex Ying}, \bibinfo{person}{Ruining He},
  \bibinfo{person}{Kaifeng Chen}, \bibinfo{person}{Pong Eksombatchai},
  \bibinfo{person}{William~L Hamilton}, {and} \bibinfo{person}{Jure Leskovec}.}
  \bibinfo{year}{2018}\natexlab{}.
\newblock \showarticletitle{Graph convolutional neural networks for web-scale
  recommender systems}. In \bibinfo{booktitle}{\emph{International Conference
  on Knowledge Discovery and Data Mining (KDD)}}. \bibinfo{pages}{974--983}.
\newblock


\bibitem[Zhan et~al\mbox{.}(2022)]%
        {zhan2022deconfounding}
\bibfield{author}{\bibinfo{person}{Ruohan Zhan}, \bibinfo{person}{Changhua
  Pei}, \bibinfo{person}{Qiang Su}, \bibinfo{person}{Jianfeng Wen},
  \bibinfo{person}{Xueliang Wang}, \bibinfo{person}{Guanyu Mu},
  \bibinfo{person}{Dong Zheng}, \bibinfo{person}{Peng Jiang}, {and}
  \bibinfo{person}{Kun Gai}.} \bibinfo{year}{2022}\natexlab{}.
\newblock \showarticletitle{Deconfounding Duration Bias in Watch-time
  Prediction for Video Recommendation}. In
  \bibinfo{booktitle}{\emph{International Conference on Knowledge Discovery and
  Data Mining (KDD)}}. \bibinfo{pages}{4472--4481}.
\newblock


\bibitem[Zhang et~al\mbox{.}(2021b)]%
        {zhang2021understanding}
\bibfield{author}{\bibinfo{person}{Fanjin Zhang}, \bibinfo{person}{Jie Tang},
  \bibinfo{person}{Xueyi Liu}, \bibinfo{person}{Zhenyu Hou},
  \bibinfo{person}{Yuxiao Dong}, \bibinfo{person}{Jing Zhang},
  \bibinfo{person}{Xiao Liu}, \bibinfo{person}{Ruobing Xie},
  \bibinfo{person}{Kai Zhuang}, \bibinfo{person}{Xu Zhang}, {et~al\mbox{.}}}
  \bibinfo{year}{2021}\natexlab{b}.
\newblock \showarticletitle{Understanding WeChat user preferences and “wow”
  diffusion}.
\newblock \bibinfo{journal}{\emph{Transactions on Knowledge and Data
  Engineering (TKDE)}} \bibinfo{volume}{34}, \bibinfo{number}{12}
  (\bibinfo{year}{2021}), \bibinfo{pages}{6033--6046}.
\newblock


\bibitem[Zhang et~al\mbox{.}(2019)]%
        {zhang2019star}
\bibfield{author}{\bibinfo{person}{Jiani Zhang}, \bibinfo{person}{Xingjian
  Shi}, \bibinfo{person}{Shenglin Zhao}, {and} \bibinfo{person}{Irwin King}.}
  \bibinfo{year}{2019}\natexlab{}.
\newblock \showarticletitle{Star-gcn: Stacked and reconstructed graph
  convolutional networks for recommender systems}.
\newblock \bibinfo{journal}{\emph{International Joint Conference on Artificial
  Intelligence (IJCAI)}} (\bibinfo{year}{2019}).
\newblock


\bibitem[Zhang et~al\mbox{.}(2022)]%
        {zhang2022dynamic}
\bibfield{author}{\bibinfo{person}{Mengqi Zhang}, \bibinfo{person}{Shu Wu},
  \bibinfo{person}{Xueli Yu}, \bibinfo{person}{Qiang Liu}, {and}
  \bibinfo{person}{Liang Wang}.} \bibinfo{year}{2022}\natexlab{}.
\newblock \showarticletitle{Dynamic graph neural networks for sequential
  recommendation}.
\newblock \bibinfo{journal}{\emph{Transactions on Knowledge and Data
  Engineering (TKDE)}} \bibinfo{volume}{35}, \bibinfo{number}{5}
  (\bibinfo{year}{2022}), \bibinfo{pages}{4741--4753}.
\newblock


\bibitem[Zhang et~al\mbox{.}(2023)]%
        {zhang2023automated}
\bibfield{author}{\bibinfo{person}{Qianru Zhang}, \bibinfo{person}{Chao Huang},
  \bibinfo{person}{Lianghao Xia}, \bibinfo{person}{Zheng Wang},
  \bibinfo{person}{Zhonghang Li}, {and} \bibinfo{person}{Siuming Yiu}.}
  \bibinfo{year}{2023}\natexlab{}.
\newblock \showarticletitle{Automated Spatio-Temporal Graph Contrastive
  Learning}. In \bibinfo{booktitle}{\emph{The Web Conference (WWW)}}.
  \bibinfo{pages}{295--305}.
\newblock


\bibitem[Zhang et~al\mbox{.}(2021a)]%
        {zhang2021motif}
\bibfield{author}{\bibinfo{person}{Zaixi Zhang}, \bibinfo{person}{Qi Liu},
  \bibinfo{person}{Hao Wang}, \bibinfo{person}{Chengqiang Lu}, {and}
  \bibinfo{person}{Chee-Kong Lee}.} \bibinfo{year}{2021}\natexlab{a}.
\newblock \showarticletitle{Motif-based graph self-supervised learning for
  molecular property prediction}. In \bibinfo{booktitle}{\emph{Advances in
  Neural Information Processing Systems (NeurIPS)}}.
  \bibinfo{pages}{15870--15882}.
\newblock


\bibitem[Zhou et~al\mbox{.}(2020)]%
        {zhou2020s3}
\bibfield{author}{\bibinfo{person}{Kun Zhou}, \bibinfo{person}{Hui Wang},
  \bibinfo{person}{Wayne~Xin Zhao}, \bibinfo{person}{Yutao Zhu},
  \bibinfo{person}{Sirui Wang}, \bibinfo{person}{Fuzheng Zhang},
  \bibinfo{person}{Zhongyuan Wang}, {and} \bibinfo{person}{Ji-Rong Wen}.}
  \bibinfo{year}{2020}\natexlab{}.
\newblock \showarticletitle{S3-rec: Self-supervised learning for sequential
  recommendation with mutual information maximization}. In
  \bibinfo{booktitle}{\emph{International Conference on Information \&
  Knowledge Management (CIKM)}}. \bibinfo{pages}{1893--1902}.
\newblock


\end{thebibliography}

\appendix

\end{document}